\documentclass[twocolumn]{aastex63}
\usepackage{graphicx}
\usepackage{wrapfig}
\usepackage{float}
\usepackage{booktabs}
\usepackage{natbib}
\usepackage{url}
\usepackage[caption=false]{subfig}

\newcommand{\PSUAA}{Department of Astronomy \& Astrophysics, 525 Davey Laboratory, The Pennsylvania State University, University Park, PA, 16802, USA}
\newcommand{\LUP}{Department of Physics, Lehigh University, Bethlehem, PA, 18015, USA}
\newcommand{\PSUCEHW}{Center for Exoplanets and Habitable Worlds, 525 Davey Laboratory, The Pennsylvania State University, University Park, PA, 16802, USA}
\newcommand{\SFSUPA}{Department of Physics and Astronomy, San Francisco State University, San Francisco, CA, 94132, USA}
\newcommand{\UCBA}{Department of Astronomy, UC Berkeley, Berkeley, CA, 94720, USA}
\newcommand{\PSETI}{Penn State Extraterrestrial Intelligence Center, 525 Davey Laboratory, The Pennsylvania State University, University Park, PA, 16802, USA}
\newcommand{\UCI}{Department of Physics \& Astronomy, UC Irvine, Irvine, CA, 92697, USA}

\graphicspath{{./}{figures/}}

\shortauthors{Baum et al.}

\begin{document}

\title{Five Decades of Chromospheric Activity in 59 Sun-like Stars and New Maunder Minimum Candidate HD 166620}


\author[0000-0002-9021-9780]{Anna C. Baum}
\affil{\PSUAA}
\affil{\LUP}

\author[0000-0001-6160-5888]{Jason T.\ Wright}
\affil{\PSUAA}
\affil{\PSUCEHW}
\affil{\PSETI}

\author[0000-0002-4927-9925]{Jacob K. Luhn}
\affil{\PSUAA}
\affil{\PSUCEHW}
\affil{\UCI}

\author[0000-0002-0531-1073]{Howard Isaacson} 
\affil{\SFSUPA}
\affil{\UCBA}

\defcitealias{Baliunas95b}{B95}
\defcitealias{Wright04}{Wright et al. 2004}


\begin{abstract}

We present five decades of chromospheric activity measurements in 59 Sun-like stars as time series. These include and extend the 35 yr of stellar chromospheric activity observations by the Mount Wilson Survey (1966--2001), and continued observations at Keck by the California Planet Search (1996--). The Mount Wilson Survey was studied closely
in 1995, and revealed periodic activity cycles similar to the Sun's 11 yr cycle.
The California Planet Search provides more than five decades of measurements, significantly improving our understanding of these stars' activity behavior. We have curated the activity measurements in order to create contiguous time series, and have classified the stellar sample according to a predetermined system. We have analyzed 29 stars with periodic cycles using the Lomb-Scargle periodogram, and present best-fit sinusoids to their activity time series. We report the best-fit periods for each cycling star, along with stellar parameters (T$_{eff}$, log(g), v$sin(i)$, etc.) for the entire sample. As a first application of these data, we offer a possible Maunder minimum candidate, HD 166620.

\end{abstract}

\keywords{ }

\section{Introduction}\label{sec:intro}

\subsection{Long-term studies of stellar magnetic activity}

The study of stellar activity has evolved from the initial discovery of the 11 yr sunspot cycle to the continued observation of hundreds of stars and their chromospheric activity, both variable and invariable. \citet[][hereafter B95]{Baliunas95b} suggested that long-term observations of activity would reveal a plethora of cycling stars.

The broad Ca II H (396.8 nm) and K (393.4 nm) spectral lines that are prominent in Sun-like stars, or, stars with Sun-like mass, radius, and temperature, and  show emission peaks in their line cores, which trace stellar activity
\citepalias{Baliunas95b, Wright04}. The strength of the emission core is correlated with the level of heating in the chromosphere by magnetic fields. A common metric of the strength of these emission cores is the $S$-value, which was originally developed by the Mount Wilson Observatory HK Project \citep{Duncan91} and is roughly proportional to the equivalent width of the emission peak. The $S$-value can be calculated by summing counts of the Ca II H and K passbands and normalizing  by the counts in the violet and red continuum bands. $S$ is given by
\begin{equation}
 S=\alpha \frac{H+K}{V+R},   
\end{equation}
where $\alpha$ is a calibration factor used to align the $S$-value with the established Mount Wilson scale \citepalias{Baliunas95b, Wright04}. Very active stars will have $S \geq$ 0.5 \citep{Duncan91}. 

Another metric for studying activity is the fraction of the observed stars' total luminosity emitted by the chromosphere in the Ca II H and K lines. This value of $R^{\prime}_{HK}$  is useful when comparing the activity level of stars of varying spectral type \citep{Noyes84}. 

There are many reasons for the continued study of chromospheric activity, considering both its causes and observable effects on stars. For example, the study of activity is important for the detection of exoplanets. Radial velocity (RV) detection has proven to be a valuable tool in the discovery of planets beyond our solar system \citep{Mayor_queloz}, and has benefited from a better understanding of stellar activity. 

There are a number of sources of systematic and star-induced error in measuring RVs. A significant contributor of error due to the star is ``jitter,” caused by inhomogeneities on the surface of a star, pulsations, and variations in the amount of convective blueshift. Jitter causes variations in measured radial velocity, and can even falsely indicate the presence of an orbiting planet \citep[][and references therein]{Wright05,Luhn20}. A primary source of jitter is stellar magnetic activity, the subject of our study. 

Long-term records of chromospheric activity allow for characterization of the multiyear timescales of magnetic fields and therefore a better understanding of their origin.  Stellar magnetic fields are induced by a magnetohydrodynamic dynamo mechanism, involving interactions between plasma flows and magnetic fields. At the surface, these fields cause sunspots, flares, faculae, and other observable features \citep{Brun15}. Studies of the Sun allow us to place constraints on the dynamo origin of stellar magnetic fields, and observations of activity in Sun-like stars can be a prime resource for additional information about stellar dynamos. 
Establishing a relationship between activity and rotation (activity cycles) plays a key role in developing knowledge on dynamos, and by extending our observations of Sun-like stars, we enhance our ability to develop a stellar dynamo model. In conjunction with asteroseismology, observations of stellar activity proxies are key to forming a more cohesive understanding of the origin of stellar magnetic fields and improving models of stellar dynamos. 

Our most well-understood star, the Sun, underwent a period known as the Maunder minimum, during which activity was very low, and perhaps constant rather than periodic \citep{Eddy76}. The search for other stars that exhibit a halt in cyclic activity is ongoing, though it is difficult with the lack of long-term records of $S$-value in other stars. While our own Sun has been studied and observed for centuries \citep{Egeland17}, our records for other stars are span, at most, multiple decades. Ongoing observations provide extended time series, allowing for the continued search for stellar Maunder minimum candidates. 

Activity monitoring was pioneered by the Mount Wilson Observatory HK Project (described in Section \ref{sec:MWP}). The success of the program led to similar surveys to study the effects and implications of chromospheric activity. 

The Solar-Stellar Spectrograph (SSS) at Lowell Observatory was directly inspired by the Mount Wilson survey. It began monitoring the Sun and many Sun-like stars in 1994 \citep{Hall07}. \citet{Henry95b} utilized the Vanderbilt/Tennessee State robotic telescope to conduct a photometric survey of 66 potentially active late-type stars. This work discovered 41 stars exhibiting flux  variability and conducted spectroscopic observations at Kitt Peak National Observatory, compiling stellar parameters as well stellar activity metrics.

\citet{Saar97} investigated the effect of chromospheric activity on low-amplitude radial velocity variations. This work provided early stellar activity insight into the now-buzzing field studying exoplanet detection. With the study of activity came the analysis of relationships between the rotational and activity cycle periods, as well as other stellar parameters. \citet{Saar99} utilized the Mount Wilson survey and an extended simple dynamo model \citep{Brandenburg98} to predict magnetic dynamo cycle periods in stars. Their high-quality photometric data were later used to investigate how Ca II emission line fluxes depend on rotation and effective temperature \citep{BohmVitense07}.

Based at Lowell and Fairbord Observatories, \citet{Lockwood07} produced 13-20 yr time series for 32 Sun-like stars, examining the relationship between photospheric and chromospheric variability. Using the SSS and the Tennessee State University Automatic Photometric Telescope, \citet{Hall09} examined the correlations between activity and variability for 28 stars over a time span of 14 years. They also identified several Maunder minimum analog candidates. We discuss the SSS more thoroughly in Section \ref{subsec:SSS}. 

The Mount Wilson $S$-values have been used for several analyses in the last 10 years. \citet{Olah09} analyzed $\sim$10 yr of photometric and Ca II H\&K observations of 20 active stars and concluded that stellar cycles are typically multiple and changing. We see similar complex cycles in our own work. More recently, they were analyzed and compared with measurements from the SSS in order to confirm the $S$-index scale placement of the Sun and lead to the accurate evaluation of $S$-values in Sun-like stars \citep{Egeland17}. \citet{Egeland17} also closely examined the offsets between HKP-1 and HKP-2 data for the Sun, a subject we  discuss in Section \ref{sec:obs}. Mount Wilson and its follow-up surveys provide a wealth of information because of the sample size and long time baseline. It is crucial for identifying long-term trends, periodic and otherwise. 


We have combined the records of two activity surveys to obtain an extensive record of stellar activity over time. The time periods previously studied were short term and do not offer the same results as long-term studies. As with the solar cycle, many stars have identifiable activity cycles of varying periods, typically about 10 yr. Our sample of Sun-like stars consists of primarily spectral type G, with some F and K stars. Their masses range between a minimum of 0.7 M$_\odot$ and an outlying maximum of about 2.3 M$_\odot$, with the majority of the sample lying between 0.7 and 1.5 M$_\odot$. Effective temperatures are contained between 4900 K and 6000 K. 
Two stars with masses around 5.0 and 6.0 M$_\odot$ are discussed in Section \ref{sec:target} as having unphysical parameters and are not considered in this description.
 As expected given our baseline, several stars exhibit decade-long activity cycles for which we have observed multiple cycles.
As surveys continue, activity cycles with much longer periods will be more easily identified. We have combined several surveys of stellar activity, and present curated long-term time series for 59 well-observed stars.

\subsection{Mount Wilson Program}
\label{sec:MWP}
The Mount Wilson program was started in 1966 by Olin Wilson to study stellar chromospheric activity, and has since become one of the most extensive records of stellar chromospheric activity. From 1966 to 1977, the project utilized the ``HKP-1" photometer: a photoelectric scanner on the coudé focus of the 100 inch telescope. Measurements continued in
1977 on the ``HKP-2" photometer: a new photomultiplier on the Cassegrain focus of the 60 inch telescope \citepalias{Baliunas95b}. Measurements on HKP-2 allowed the increase in both sample size and frequency of observations. \citetalias{Baliunas95b} estimated the long-term precision of the measurements to be 1.2\%, later verified by Richard Radick\citep{HKPReadMe} to be 1-2\%. The sample size of stars continuously increased to the current record of almost 2300 stars, 35 of which were observed through 2001 \citep{HK1995, HK2001}.

In search of periodic activity cycles, \citetalias{Baliunas95b} examined chromospheric activity of 111 stars from 1966 to 1995, utilizing the data from the Mount Wilson survey \citep{Wilson78,Vaughan78,Duncan91}. They analyzed the comprehensive time series of measured $S$-values for each of these stars, identifying periodic activity variations in a significant portion of the sample. Stars with irregular or no variation in activity level were also identified. Some stellar activity candidate cycles could not be verified because an observation time equivalent to two full periods is necessary for confirmation. Periodograms were used to approximate the periods of the ``cycling" stars, and the time series of each star was presented, along with each star's annually averaged mean $S$-value, color index, and spectral type.

\citetalias{Baliunas95b} concluded that chromospheric activity depended heavily on stellar mass and age, and follows an evolutionary time scale. Activity was also noted to increase with $B$-$V$, primarily because $S$-value is sensitive to photospheric temperature. Longer intervals of observation can reveal cycles of stars with longer periods, as well as possible Maunder minima or inconsistent periodicity.

\subsection{California Planet Search}
The California Planet Search (CPS) and its predecessor surveys compose a modern program based at the Keck and Lick Observatories that primarily targets stars thought to be good planet-search targets. It captures some of the brightest, oldest F, G, and K dwarfs visible from Keck, and notably omits very active stars. 
CPS uses the High Resolution Echelle Spectrometer (HIRES) at Keck Observatory \citep{Vogt94} and the Hamilton spectrograph \citep{Vogt87} with the Shane 3 m telescope and the 0.6 m Coude Auxilary Telescope at Lick Observatory, both echelle spectrometers with high resolution \citep{Wright04}. The Keck Observatory uses an image rotor to align the slit axis with the elevation axis to keep the efficiency of measurements high. Our work includes only the measurements taken at Keck because of its lower measurement uncertainties.

\citet{Wright04} and \citet{Isaacson10} published CPS $S$-value activity catalogs, representing distinct data-analysis efforts on data taken with two different HIRES detectors. After 2004, the HIRES spectrometer was upgraded with a new CCD to a achieve an RV precision of 1 ms$^{-1}$. We label the pre-2004 detector ``HIRES-1" and the post-2004 upgraded detector ``HIRES-2."
\citet{Wright04} recorded the $S$-values of about 700 stars, and \citet{Isaacson10} recorded the $S$-values of more than 2600 stars. In addition to these two sets of published data, we also present new $S$-value measurements taken at HIRES since the publication of \citet{Isaacson10}, using the same methods described in that work, creating an even more complete record of activity measurements. 

\citet{Wright04b} proposed several conclusions on examination of long-term activity observations. Wright suggested that the low-activity and flat stars in the sample analyzed were primarily composed of subgiants; we examine this conclusion in the new, extended data. \citet{Wright04b} was particularly concerned with the (mis)identification by \citet{Baliunas90} of low-activity stars as Maunder minimum candidates on the basis of their low activity levels.  We use the extended time series to explore Maunder minimum-like events.


\section{Data} \label{sec:obs}

\subsection{$S$-Values} 
We have collected all observations of $S$-values from four sources in order to assemble the most complete time series possible from 1966-present. This includes the compilation and cross-referencing of each data set from various resources. Once combined, we addressed and corrected offsets between data sets.

\subsubsection{Mount Wilson}
The data from the Mount Wilson program for ~2300 stars were provided by \citet{HK1995} and \citet{HK2001} as found on the Harvard Dataverse. We collected the data for all stars observed between 1966 and 1995 along with the observations of 35 stars through 2001 appended. 
We note the change from HKP-1 to HKP-2 is noted in the plots and our data tables. The observations made pre-1977, in some cases, showed clear, large offsets with the more frequent post-1977 observations, most notably for HD 22072, HD 23249, and HD 217014. We remedied these offsets by shifting the median of the pre-1977 measurements to align with the median of the more extensive post-1977 measurements. 


\subsubsection{Wright et al. 2004}
We take HIRES-1 data from 1996 to 2004 from \citet{Wright04}.  For three stars (HD 3795, 10145, and 65583), we find significantly and consistently offset $S$-values between \citet{Wright04} and Mount Wilson data. Where we felt we could confidently correct these offsets, we adjusted the \citet{Wright04} data with a constant shift to have a consistent median $S$-value with the Mount Wilson data set, the most reliable resource we have on the thorough and consistent observation of $S$-values. These offsets can be explained by the calibration uncertainty between the very different instruments.

A couple of stars (HD 34411, 142267) show possible large offset due to calibration, but we chose not to correct them because the more recent data are sparse enough that the difference could be a real change in the star's activity level.

\subsubsection{Isaacson \& Fischer 2010}
We take HIRES-2 data from 2004 to 2010 from \citet{Isaacson10}. These data are the more precise of the CPS data sets. Fourteen stars in the post-2010 CPS data showed a clear offset, much like those seen in the \citet{Wright04} data, and were shifted accordingly, in the same manner.

\begin{figure}[h]
\figurenum{1}
\includegraphics[width=\linewidth]{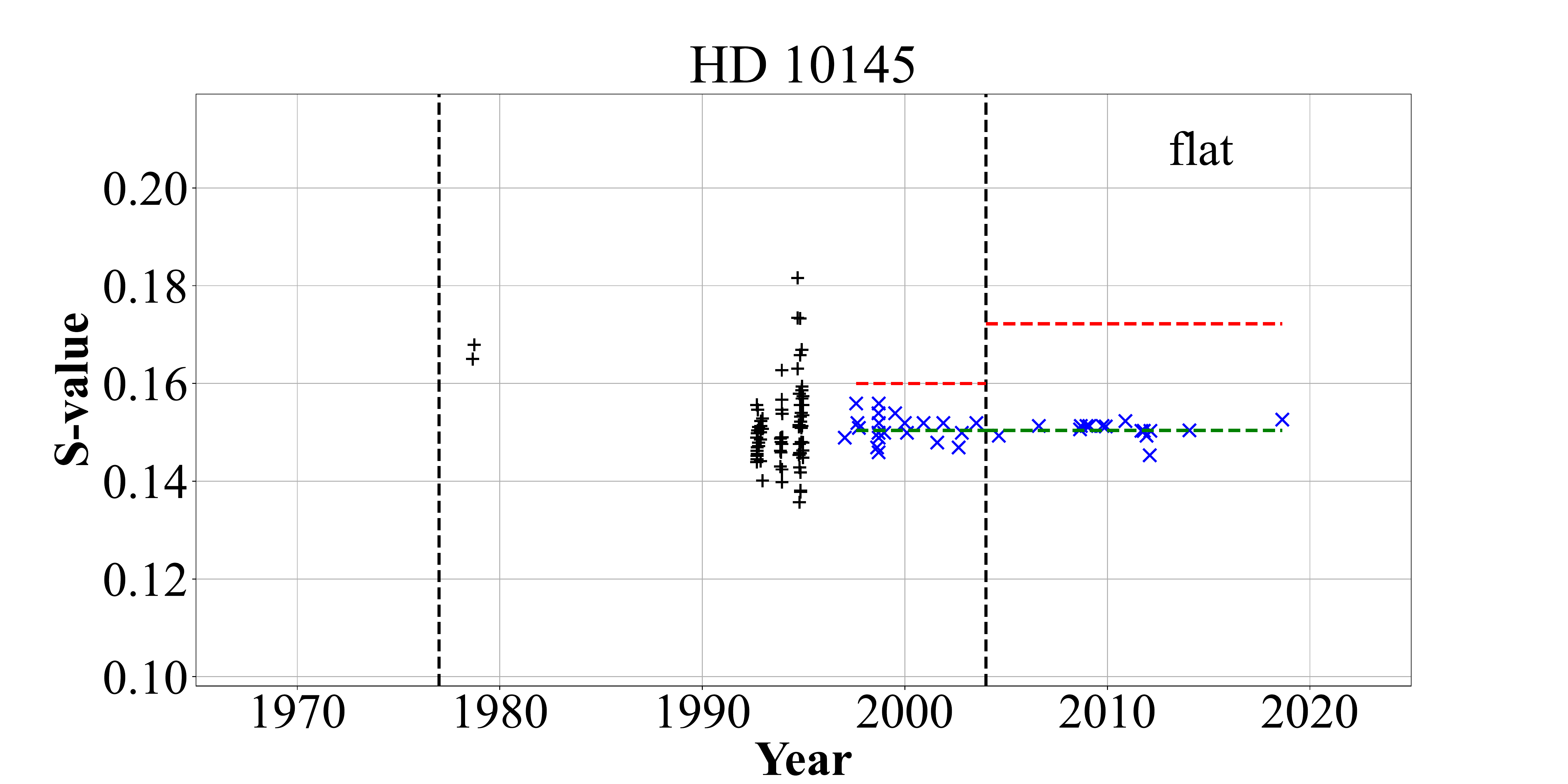}
\caption{Initial plots of HD 10145 showed large, consistent offsets in $S$-value between pre- and post-2004 CPS data. We present data with adjusted median $S$-value.}
\label{fig:1}
\end{figure}

In the individual stars' plots, we display two dashed lines when offsets are applied: one red, showing the original median of the data, and one green, showing the corrected median. HD 10145 is shown in Figure \ref{fig:1} as an example of this formatting.

\subsubsection{Post-2010 CPS Data}
CPS data dating from 2010 to 2020, serving as an extension of the \citet{Isaacson10} data, are included in these measurements. 
CPS data from 2010 to present are calculated in the same 
manner as data prior to 2010, allowing for continuity in each star's values from 2005 to present. 
Note that the use of the C2 decker to observe bright stars in  twilight and also faint stars through the night causes 
occasional issues with the $S$-values when seeing
is poor and the observation is taken at high air mass. This is 
addressed in Section \ref{subsec:seeing}.

\subsection{Stellar Parameters}

\citet{Brewer16} provides a source for precise stellar parameters for F, G, and K stars that were unavailable to \citetalias{Baliunas95b}. These parameters allowed us to retrieve effective temperature, rather than color index, for the future examination of trends in chromospheric activity, as well as surface gravity, rotational velocity, metallicity, mass, and age. All spectral properties were retrieved from the Keck HIRES spectrometer as part of the CPS. We use [Fe/H] to generalize metallicity; other abundances can be found in \citet{Brewer16}. 

Most spectra from \citet{Brewer16} had a high signal-to-noise ratio (S/N), and therefore made  observational uncertainties a negligible contribution to error in these parameters. \citet{Brewer16} quantified and minimized other sources of error, such as modeling inconsistencies or limitations, to their best ability. Because spectral type was not
included in the \citet{Brewer16} parameters, the spectral types of all stars
in our sample were retrieved from the SIMBAD database \citep{SIMBAD}.

\subsection{Target Selection}\label{sec:target}

We cross-checked the sample of 2326 stars from the Mount Wilson data with the \citet{Brewer16} data set to ensure each star had precise measurements of stellar parameters recorded, which limited the sample to 189 stars.   

Four stars, including HD 178911B, HD 145958A and 145958B, and HD 219834B were noted to have unphysical parameters relating to either mass (recorded masses greater than 3 M$_{\odot}$) or age (stellar ages greater than the age of the universe). These results were likely due to errors in spectroscopic analysis. We have decided to include these stars regardless, in order to present a more complete catalog of chromospheric activity in stars.

We combine CPS observations with the Mount
Wilson records, providing a far more extensive record of activity up to 2020. The time series spans 1966-present, with some gaps but overall enough data to properly observe trends in activity over a long time period. 

From our sample of 189 stars with Mount Wilson observations and measured stellar parameters, 59 stars have significant and classifiable records of $S$-value from both Mount Wilson and CPS. Of the 59 stars included in our sample, 26 were also evaluated in \citetalias{Baliunas95b}.

\subsection{Outliers}
For the entire length of combined data, single data point outliers were removed in order
to improve analysis of long-term trends. Any data points points more than three standard deviations from the mean were
removed. We checked to confirm that this process did not truncate  natural variations of stars due to long-term activity trends. 

In addition to the occasional instrumental and data reduction
errors, our sigma-clipping outlier rejection may have removed possible activity peaks or stellar flares. However, we are interested in the curated time series to identify low-frequency long-term changes in activity, so the rejection of flares is not of concern to us. We include all points rejected due to sigma-clipping in the supplemental data with flags labeled `sigma' to indicate their removal. 

Other $S$-values that were possibly unphysical or due to instrumental or data reduction error left behind after the sigma-clipping procedure were removed by hand. This was limited to stars HD 140144, HD 1461, and HD 182572, each of which had single instance observations of $S$-values measurements two or more standard deviations from the median $S$-value. For the case of HD 182572, we also note a high density of points at an $S$-value around 3$\sigma$ from the mean. This particular instance could be followed up in future work as a possible activity burst. For the purposes of long-term trend analysis, these points are left out of our analysis and can be found in supplemental data.

Several stars have instances of recorded $S$-values of 0.00, which is unphysical, and is indicative of bad data, most likely resulting from instrumental or calibration errors. These were also removed from the data set. All points removed due to certain instrumental or calibration error, such as those with 0.00 $S$-values or measurements taken with the C2 decker under poor seeing conditions, discussed in Section \ref{subsec:seeing}, are excluded from the final data sample entirely, including supplemental material, because they are of no use for future analysis.

\subsection{Seeing}\label{subsec:seeing}
In the time series, `dips' in $S$-value were found to consistently occur in post-2004 CPS data, with implementation of the C2 decker after 2009. Similar to the offsets, these dips were initially noted in our initial examination of the time series. We referenced the original observation log sheets in search of possible causes for these dips. Using the decker and seeing from CPS observation records, we determined that the combination of the C2 decker and poor seeing conditions was correlated to dipping measurements of $S$-values. We analyzed the relationship between seeing and $S$-values for observations made using the C2 decker, and found the drop off into dipping $S$-values occurred at a seeing of 1.5", and progressively declined as seeing increased. 

In the blue CCD, which contains the Ca II H\&K lines, the spectral orders are closer together than in the middle and red CCDs. This results in overlapping orders when the C2 decker is used and the seeing is $>$ 1.5". Contamination from adjacent orders makes the $S$-value measurement unreliable. Therefore, CPS data observations made using the C2 decker with seeing greater than 1.5" were removed. These adjustments made possible the curated, contiguous time series presented here. 


\subsection{Final Sample}
Stars without considerable added value were also removed from our final sample. This included any stars with few observations from either the Mount Wilson or CPS program. Of the 189 stars, we deem 59 stars to include significantly improved data from the results presented in \citetalias{Baliunas95b}. The final sample of stars is fully represented in Table \ref{startable}. 

Table \ref{svaluetable} shows the first five lines of the file containing HD number, $S$-value observation, Gregorian date of observation, and instrument used. The complete table contains the curated observations of all 59 stars in the sample. We report many significant figures in our table; however the actual times of observation are all only given to precision one day for HKP-1 and HKP-2 1977-1979 configuration, up to 0.0001 days for subsequent HKP-2 configurations, 0.0001 days for HIRES-1, and 0.01 days for HIRES-2. Observations taken with HKP-1 and early HKP-2 are given arbitrary or inaccurate times of observation, which correlate to times the star was at very low altitude. We include flags on entries for which precision is only within one day. For data entries including two flags--one day precision and sigma-clipped, we separate them by a forward slash, i.e., `1day/sigma'. 



\begin{deluxetable}{ccccc}[H]
\tablecaption{Curated $S$-values}
\tablehead{\colhead{Star} & \colhead{$S$-value} & \colhead{Time} & \colhead{Instrument} & \colhead{Flags}}
\startdata
1388 & 0.1533 & 1992.67864290 & HKP-2 & none \\
1388 & 0.1463 & 1992.67864822 & HKP-2 & none \\
1388 & 0.1504 & 1992.67865224 & HKP-2 & none \\
1388 & 0.1552 & 1992.69221205 & HKP-2 & none \\
1388 & 0.1552 & 1992.69221738 & HKP-2 & none
\enddata
\tablecomments{This paper shows only a sample of the full table. The complete table is available electronically.}
\end{deluxetable}
\label{svaluetable}%

\section{Analysis} \label{sec:anal}


Prior to analyzing each time series, we calculated the cadence and baseline of observations for
each star. The cadence is denoted by the average number of observations per year, and baseline is the number of years the star was observed. For stars with gaps in the records, we determined the baseline using the earliest and latest observations recorded. We then calculated and recorded the median $S$-value over the total baseline in Table \ref{startable}. We separated main-sequence stars and subgiants by surface gravity, defining main-sequence stars as those with log$(g)>4.2$ and subgiants as stars with log$(g)\leq 4.2$.  


Initial assessment of the $S$-value over time and was done by eye using specific criteria, loosely based on those used in \citetalias{Baliunas95b}. To identify a star as cycling, it must have:
\begin{enumerate}
    \item clear periodic variation
    \item a significant number of observations
    \item at least two full periods (preferably)
    \item similar and consistent amplitude peaks and troughs (otherwise labeled as insuf or var)
\end{enumerate}

\begin{figure}[ht]
\figurenum{2}
\centering
\includegraphics[width=\linewidth]{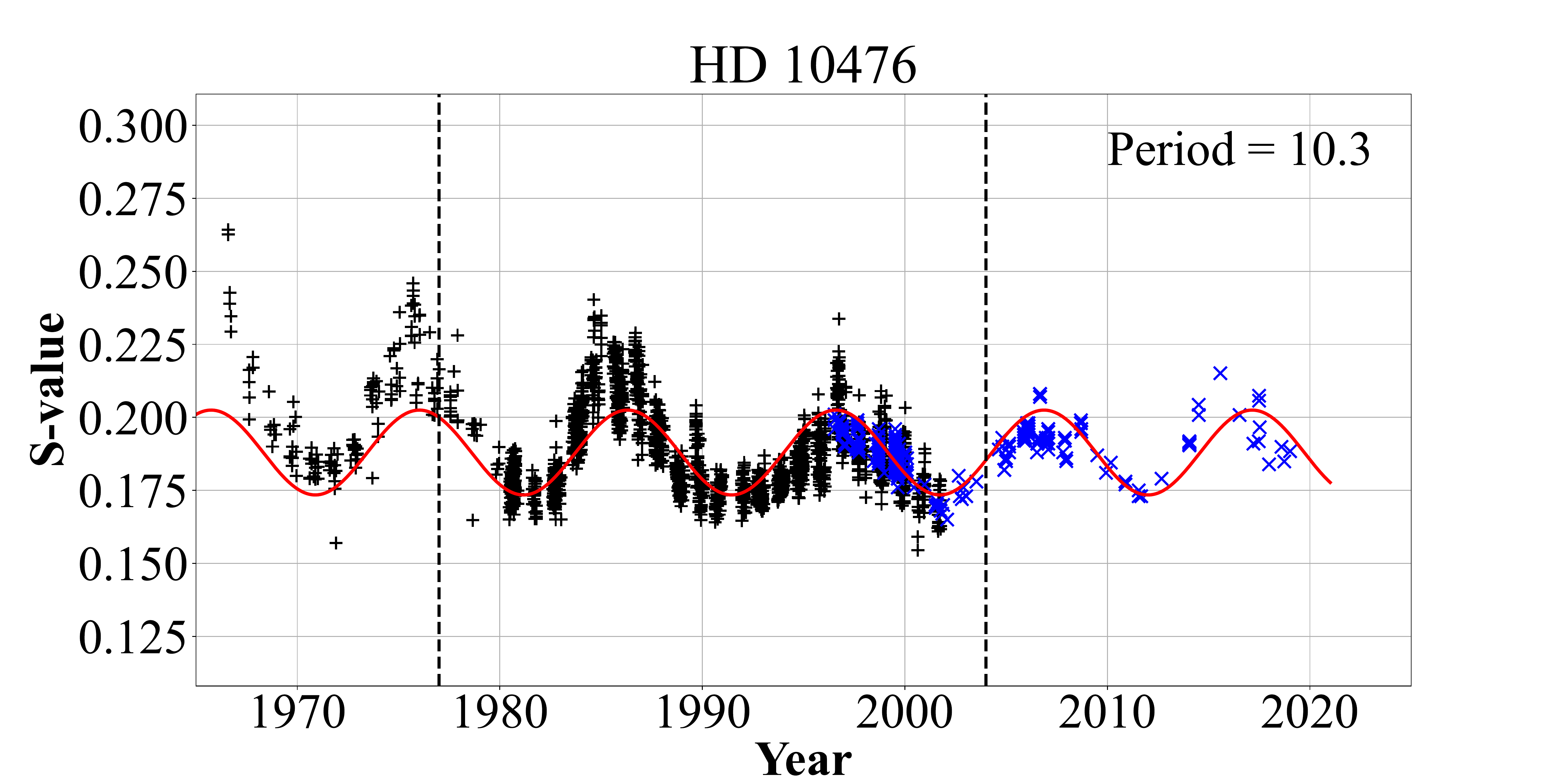}
\caption{Activity vs. time in 59 Sun-like stars: the complete figure set (59 images) is available in the online journal.}
\label{fig:2}
\end{figure}

Other time series classifications included ``flat" (no observed variation in activity),
``long" (possible long-term cycle), ``var" (clear, nonperiodic variations), and
``insuf" (insufficient data to make a classification). ``Insuf" was a heavily
used classification, in order to factor out any stars where classification was
unclear. The criteria for these classifications were based on looser foundations than those from \citetalias{Baliunas95b}, because our primary goal was more focused on data presentation, and presenting curated data for public use. \figsetstart
\figsetnum{2}
\figsettitle{Activity Time Series}
\figsetgrpstart
\figsetgrpnum{2.1}
\figsetgrptitle{1388}
\figsetplot{1388.pdf}
\figsetgrpnote{Activity vs time plot for HD 1388.}
\figsetgrpend
\figsetgrpstart
\figsetgrpnum{2.2}
\figsetgrptitle{1461}
\figsetplot{1461.pdf}
\figsetgrpnote{Activity vs time plot for HD 1461.}
\figsetgrpend
\figsetgrpstart
\figsetgrpnum{2.3}
\figsetgrptitle{3795}
\figsetplot{3795.pdf}
\figsetgrpnote{Activity vs time plot for HD 3795. HIRES-1 and HIRES-2 data are offset-corrected.}
\figsetgrpend
\figsetgrpstart
\figsetgrpnum{2.4}
\figsetgrptitle{4307}
\figsetplot{4307.pdf}
\figsetgrpnote{Activity vs time plot for HD 4307.}
\figsetgrpend
\figsetgrpstart
\figsetgrpnum{2.5}
\figsetgrptitle{4628}
\figsetplot{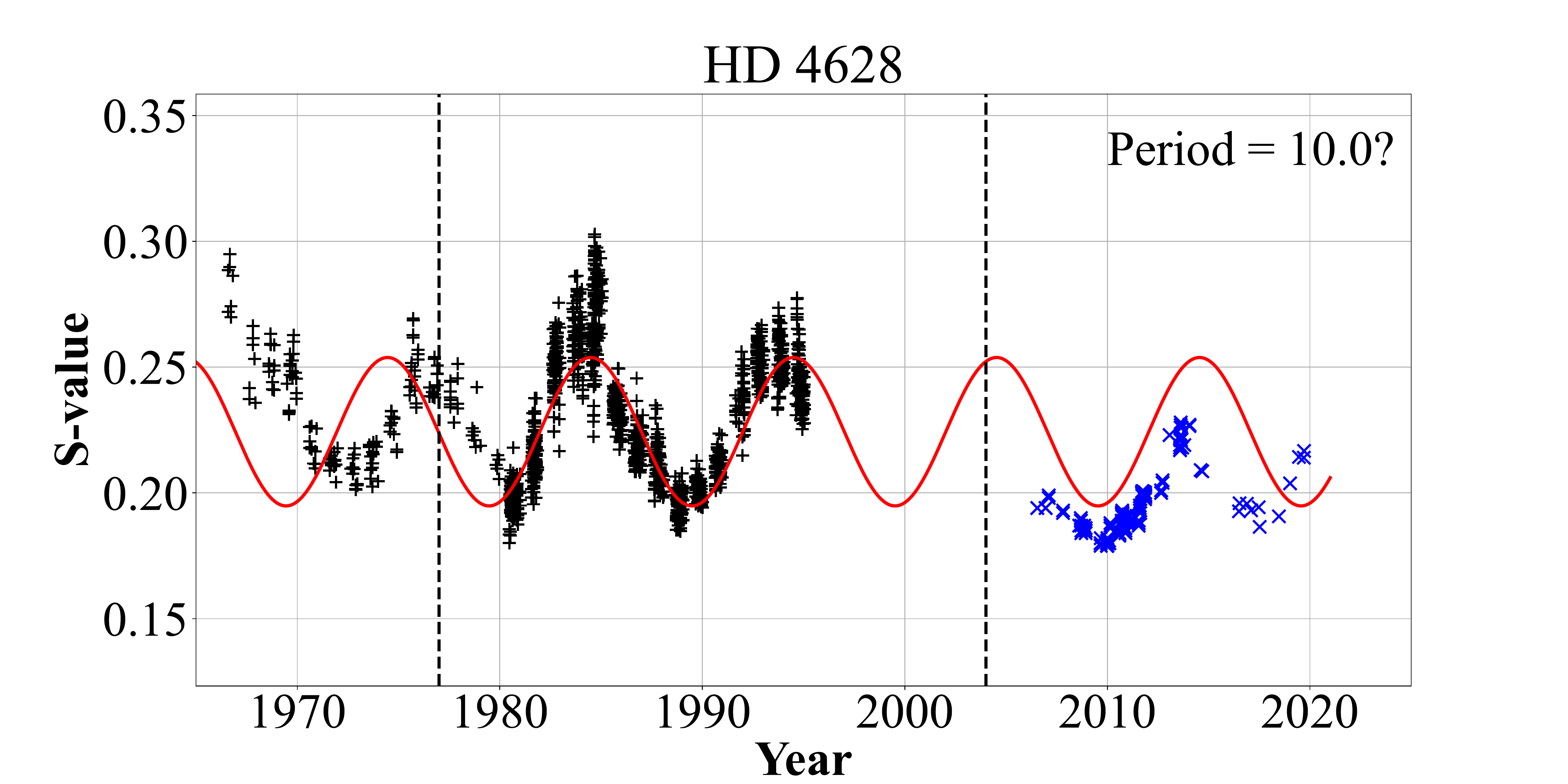}
\figsetgrpnote{Activity vs time plot for HD 4628. Cycling in HD 4628 has an inconsistent phase.}
\figsetgrpend
\figsetgrpstart
\figsetgrpnum{2.6}
\figsetgrptitle{7924}
\figsetplot{7924.pdf}
\figsetgrpnote{Activity vs time plot for HD 7924.}
\figsetgrpend
\figsetgrpstart
\figsetgrpnum{2.7}
\figsetgrptitle{9562}
\figsetplot{9562.pdf}
\figsetgrpnote{Activity vs time plot for HD 9562. HIRES-2 data is offset-corrected.}
\figsetgrpend
\figsetgrpstart
\figsetgrpnum{2.8}
\figsetgrptitle{10145}
\figsetplot{10145.pdf}
\figsetgrpnote{Activity vs time plot for HD 10145.}
\figsetgrpend
\figsetgrpstart
\figsetgrpnum{2.9}
\figsetgrptitle{10476}
\figsetplot{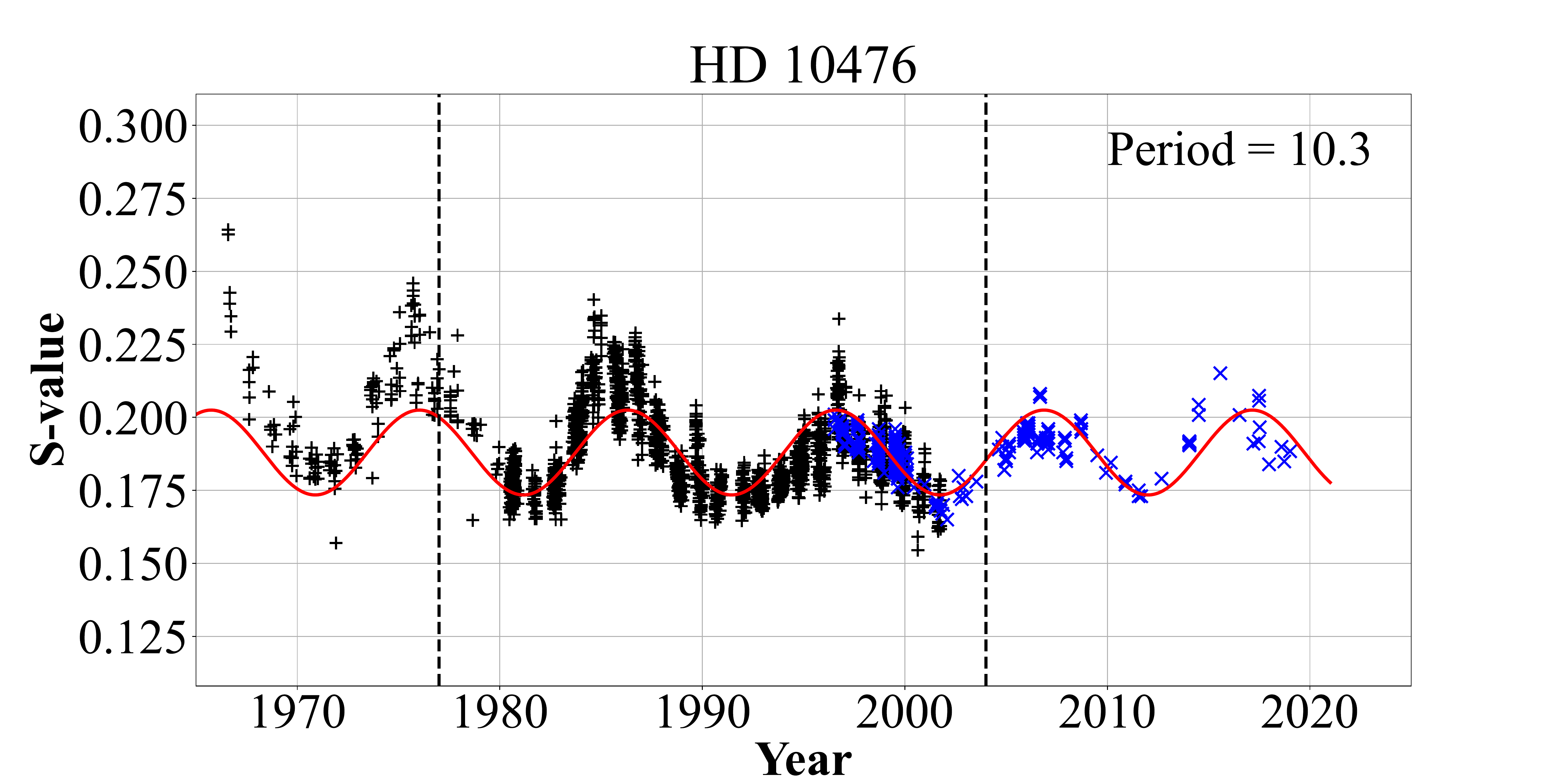}
\figsetgrpnote{Activity vs time plot for HD 10476.}
\figsetgrpend
\figsetgrpstart
\figsetgrpnum{2.22}
\figsetgrptitle{50692}
\figsetplot{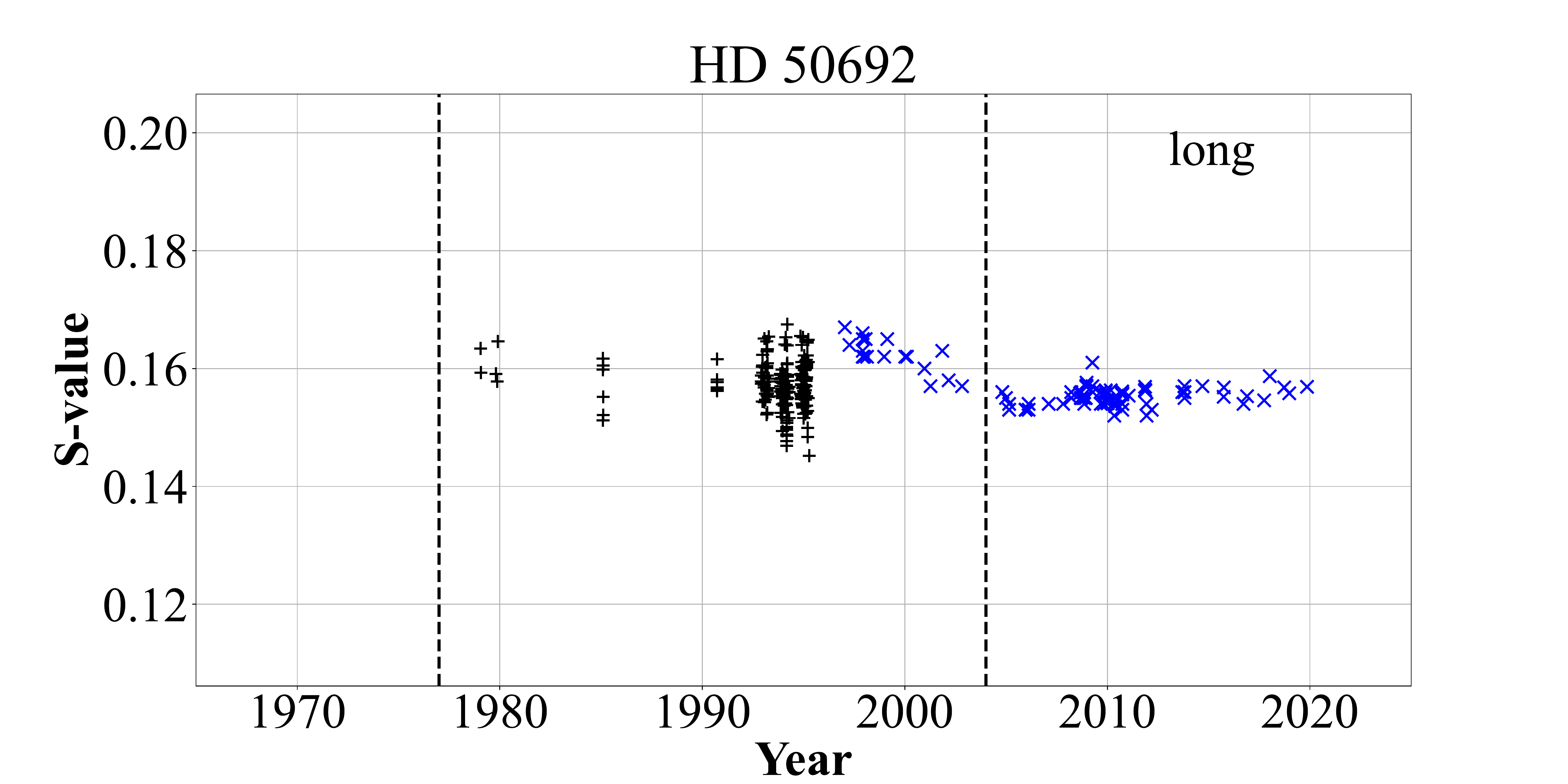}
\figsetgrpnote{Activity vs time plot for HD 50692.}
\figsetgrpend
\figsetgrpstart
\figsetgrpnum{2.23}
\figsetgrptitle{52711}
\figsetplot{52711.pdf}
\figsetgrpnote{Activity vs time plot for HD 52711.}
\figsetgrpend
\figsetgrpstart
\figsetgrpnum{2.24}
\figsetgrptitle{65583}
\figsetplot{65583.pdf}
\figsetgrpnote{Activity vs time plot for HD 65583. HIRES-1 data is offset-corrected.}
\figsetgrpend
\figsetgrpstart
\figsetgrpnum{2.25}
\figsetgrptitle{72905}
\figsetplot{72905.pdf}
\figsetgrpnote{Activity vs time plot for HD 72905. HIRES-2 data is offset-corrected.}
\figsetgrpend
\figsetgrpstart
\figsetgrpnum{2.26}
\figsetgrptitle{75732}
\figsetplot{75732.pdf}
\figsetgrpnote{Activity vs time plot for HD 75732.}
\figsetgrpend
\figsetgrpstart
\figsetgrpnum{2.27}
\figsetgrptitle{86728}
\figsetplot{86728.pdf}
\figsetgrpnote{Activity vs time plot for HD 86728.}
\figsetgrpend
\figsetgrpstart
\figsetgrpnum{2.28}
\figsetgrptitle{100180}
\figsetplot{100180.pdf}
\figsetgrpnote{Activity vs time plot for HD 100180.}
\figsetgrpend
\figsetgrpstart
\figsetgrpnum{2.29}
\figsetgrptitle{101501}
\figsetplot{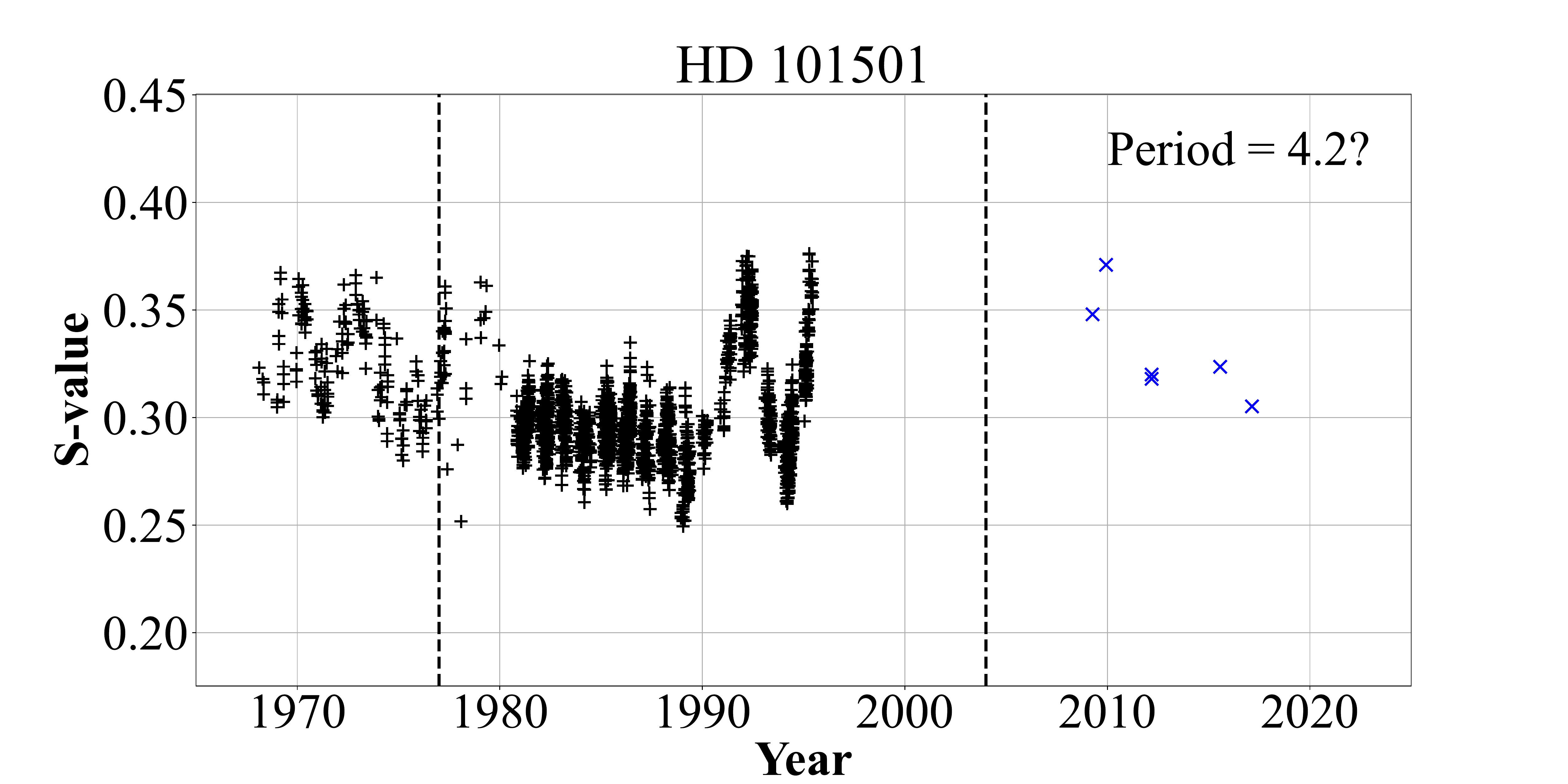}
\figsetgrpnote{Activity vs time plot for HD 101501. HD 101501 exhibits high variability preceding and following 10 years of low variablility.}
\figsetgrpend
\figsetgrpstart
\figsetgrpnum{2.30}
\figsetgrptitle{115617}
\figsetplot{115617.pdf}
\figsetgrpnote{Activity vs time plot for HD 115617.}
\figsetgrpend
\figsetgrpstart
\figsetgrpnum{2.31}
\figsetgrptitle{126053}
\figsetplot{126053.pdf}
\figsetgrpnote{Activity vs time plot for HD 126053.}
\figsetgrpend
\figsetgrpstart
\figsetgrpnum{2.32}
\figsetgrptitle{132142}
\figsetplot{132142.pdf}
\figsetgrpnote{Activity vs time plot for HD 132142. HIRES-2 data is offset-corrected.}
\figsetgrpend
\figsetgrpstart
\figsetgrpnum{2.33}
\figsetgrptitle{141004}
\figsetplot{141004.pdf}
\figsetgrpnote{Activity vs time plot for HD 141004.}
\figsetgrpend
\figsetgrpstart
\figsetgrpnum{2.34}
\figsetgrptitle{142267}
\figsetplot{142267.pdf}
\figsetgrpnote{Activity vs time plot for HD 142267.}
\figsetgrpend
\figsetgrpstart
\figsetgrpnum{2.35}
\figsetgrptitle{143761}
\figsetplot{143761.pdf}
\figsetgrpnote{Activity vs time plot for HD 143761.}
\figsetgrpend
\figsetgrpstart
\figsetgrpnum{2.36}
\figsetgrptitle{145958A}
\figsetplot{145958A.pdf}
\figsetgrpnote{Activity vs time plot for HD 145958A.}
\figsetgrpend
\figsetgrpstart
\figsetgrpnum{2.37}
\figsetgrptitle{145958B}
\figsetplot{145958B.pdf}
\figsetgrpnote{Activity vs time plot for HD 145958B.}
\figsetgrpend
\figsetgrpstart
\figsetgrpnum{2.38}
\figsetgrptitle{146233}
\figsetplot{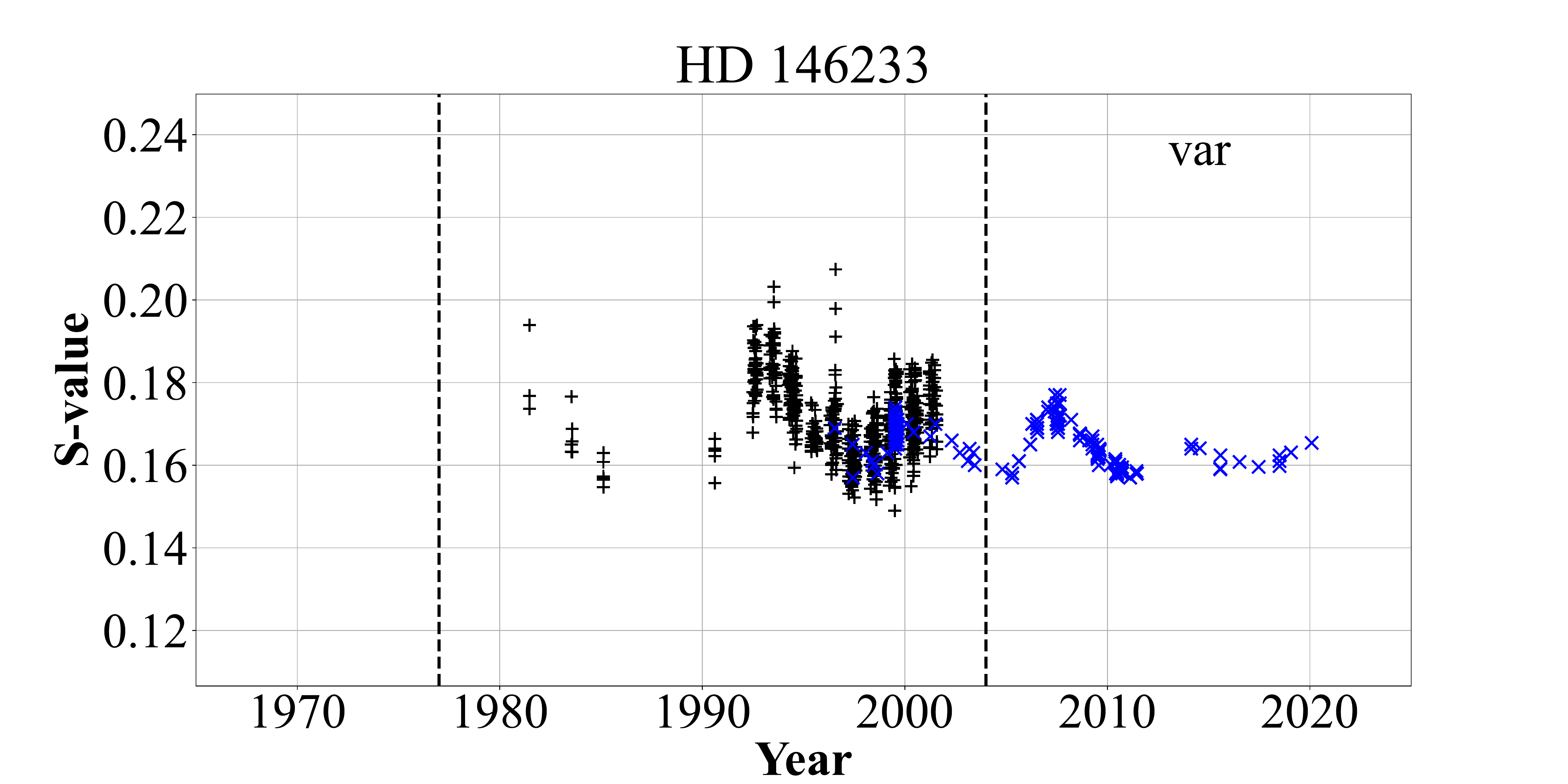}
\figsetgrpnote{Activity vs time plot for HD 146233. Cycling in HD 146233 follows a highly variable trend.}
\figsetgrpend
\figsetgrpstart
\figsetgrpnum{2.39}
\figsetgrptitle{152391}
\figsetplot{152391.pdf}
\figsetgrpnote{Activity vs time plot for HD 152391. HIRES-2 data is offset-corrected.}
\figsetgrpend
\figsetgrpstart
\figsetgrpnum{2.40}
\figsetgrptitle{157214}
\figsetplot{157214.pdf}
\figsetgrpnote{Activity vs time plot for HD 157214.}
\figsetgrpend
\figsetgrpstart
\figsetgrpnum{2.41}
\figsetgrptitle{159222}
\figsetplot{159222.pdf}
\figsetgrpnote{Activity vs time plot for HD 159222.}
\figsetgrpend
\figsetgrpstart
\figsetgrpnum{2.42}
\figsetgrptitle{166620}
\figsetplot{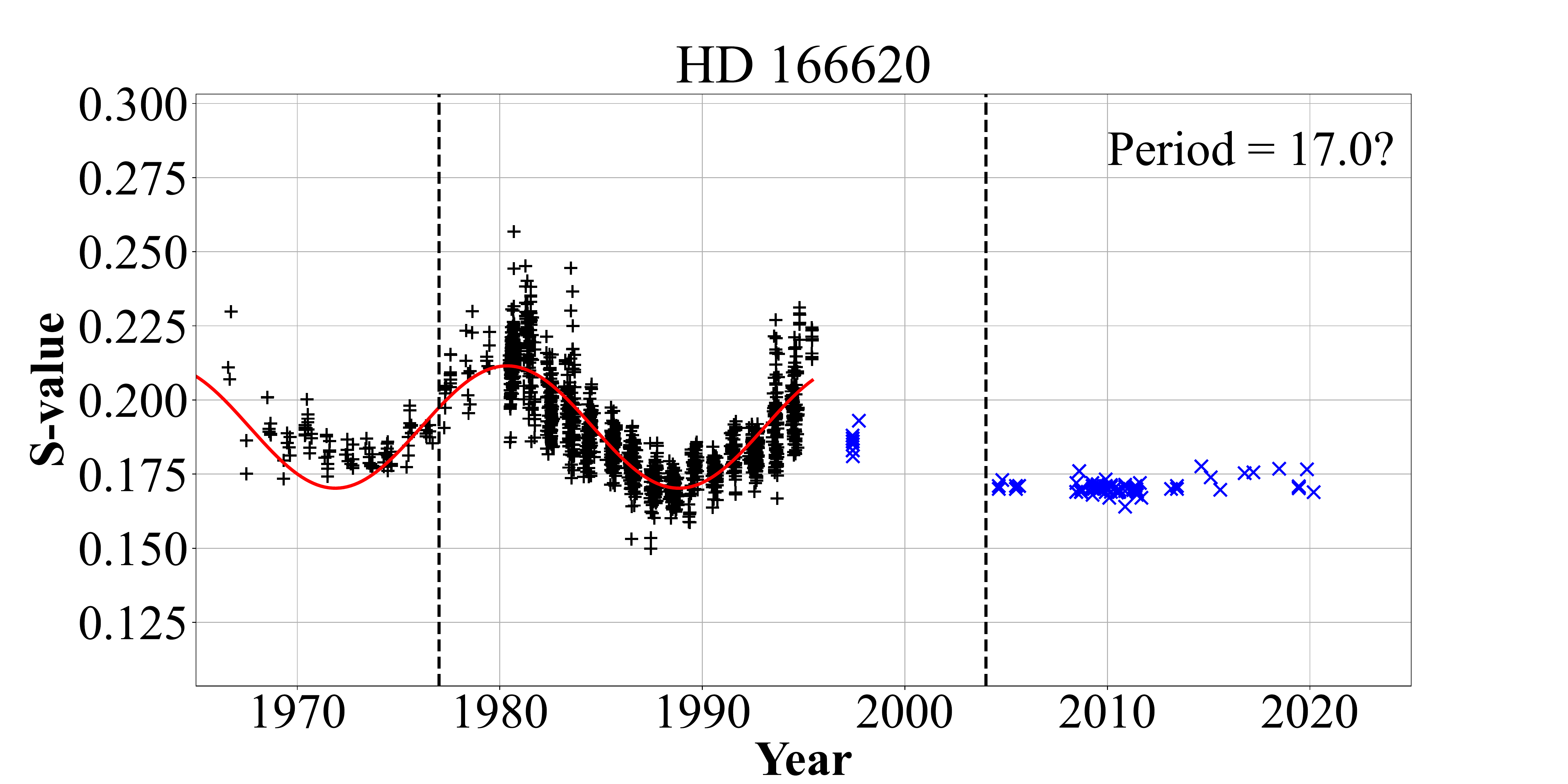}
\figsetgrpnote{Activity vs time plot for HD 166620. Cycling in HD 166620 appears to cease between 1995 and 2004. HD 166620 is proposed to be a Maunder Minimum Candidate according to the observed behavior.}
\figsetgrpend
\figsetend\label{timeseries}

\subsection{Periodograms} \label{subsec:pergram}
\begin{figure}[ht]%
\figurenum{3}
\centering
\subfloat[Cycling in HD 4628 has a phase inconsistent with our sinusoidal model.]
{\includegraphics[width=\columnwidth]{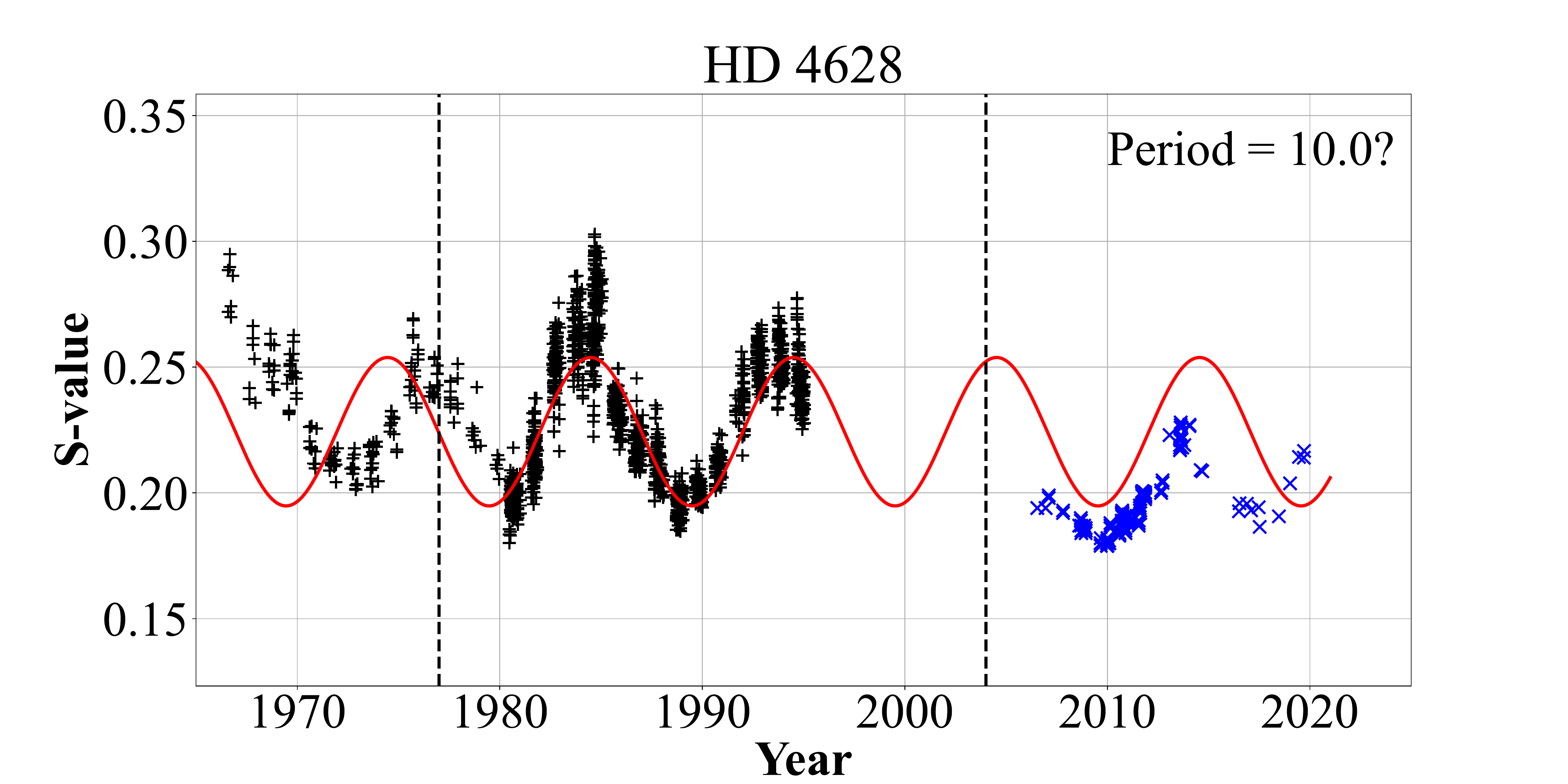}}
\\
\subfloat[Cycling in HD 146233 follows a highly variable trend.]{\includegraphics[width=\columnwidth]{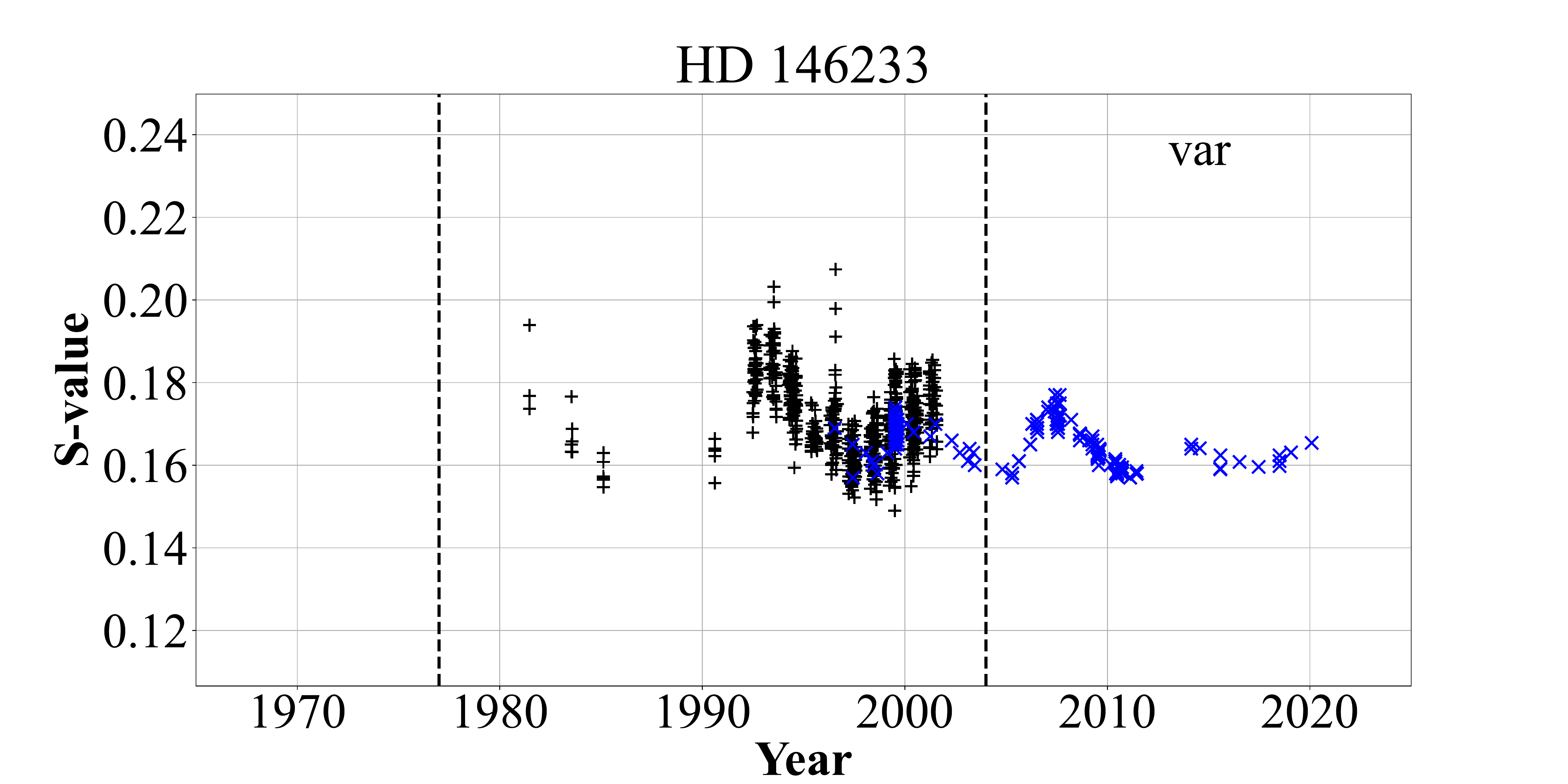}}
\caption{}
\label{fig:varcycles}
\end{figure}

Similar to  \citetalias{Baliunas95b}, we next calculated a periodogram \citep{Lomb76,Scargle82} for each time series to assist in classification.
We found a period estimate for each cycling star as that which had the most power in the periodogram. The period was used to model a sinusoid  with the same period and approximate amplitude and phase to best fit and illustrate the activity cycle. A good example of a well-fit cycle is HD 10476 in Figure \ref{fig:2} We chose this method to generate a simple description of the apparent activity cycle, but activity cycles are not strictly periodic (see Figure \ref{fig:varcycles}), and certainly not sinusoidal. 

\begin{figure}[ht]%
\figurenum{4}
\centering
\subfloat[HD 166620 appears to have entered a Maunder minimum between its final observations with HIRES-1 and first observations with HIRES-2. ]{%
    \includegraphics[width=\linewidth]{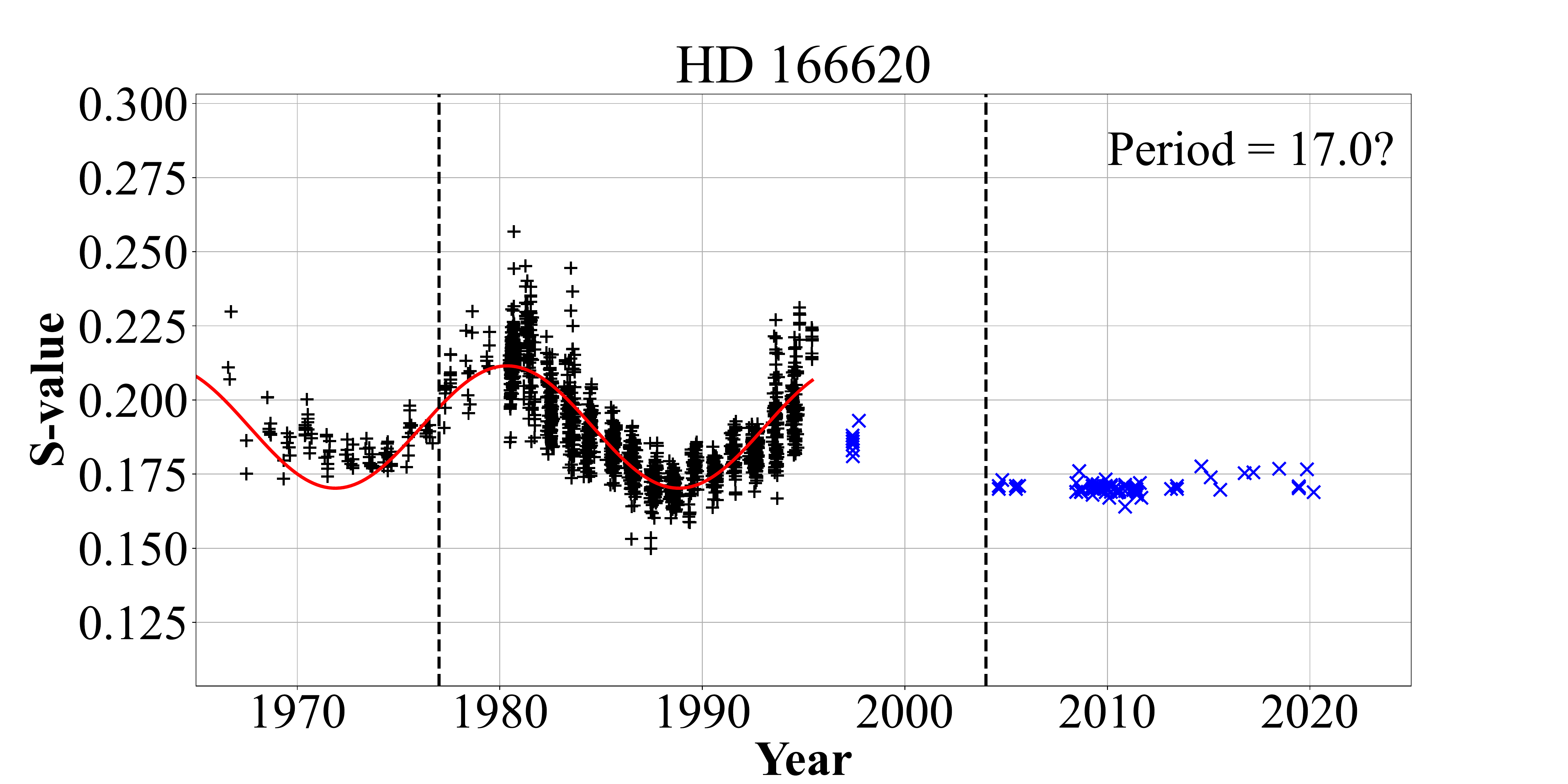}%
}

\subfloat[HD 101501 experienced 10 yr of lower activity, a much lower amplitude cycle than the rest of its cycle.]{
    \includegraphics[width=\linewidth]{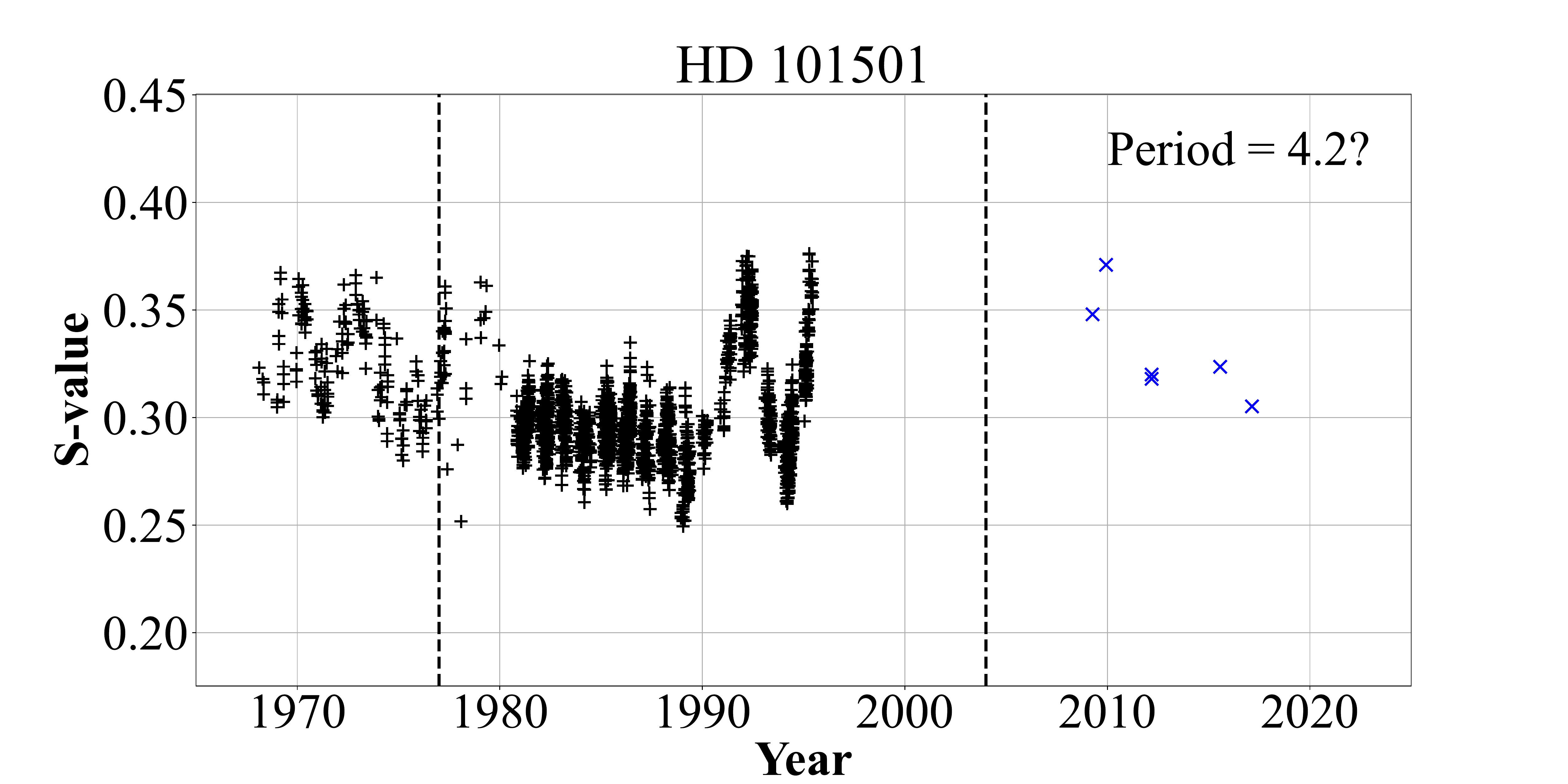}%
}
\caption{}
\label{fig:maundmin}
\end{figure}

Some stars, such as HD 4628, 101501, 146233, 166620, and 190406, have variations in their cycle or large gaps in data collection that result in an unreliable periodogram analysis. Two of these stars, HD 101501 and 166620, are discussed in greater detail in section \ref{subsec:mmcan}.
For these stars, which appear to be cycling, as seen in Figure \ref{fig:maundmin}, a period estimate was made by eye in order to produce a curve of best fit. HD 190406 appears to have a secondary periodic trend, to which we did not fit a curve. Additionally, stars with very low-amplitude variability were not closely analyzed for periodic trends, and are classified as either `flat?' or variable. All period estimates for stars with cycling activity are listed in Table \ref{table:startable}. Stars for which estimates were made by eye rather than periodogram include a `?' with the value. In accordance with \citetalias{Baliunas95b}, periods are estimated only to one-10th of a year and are not proposed as precise.

We verify the estimated periods of each star using a sinusoidal model. This could not always produce a good fit because stellar activity variation is not strictly periodic; 
stars cycle with a far more complicated variance and are not consistent in period over many cycles. Some stars could be entering or exiting a Maunder minimum, discussed in section \ref{subsec:mmcan}, or could have a variance that appears to be cyclic but is not.
Additionally, many of the stars appear to have a cycle that changes 
over time. 
\citet{Olah09} examined this behavior in detail. HD 4628 and 146233 are two good examples of this phenomenon.




\section{Results} \label{sec:results}

We present a table of the 59 stars in our sample along with their classification and several other parameters. We also present the $\sim$ 50 yr time series for these stars, doubling the length of time series previously presented by \citetalias{Baliunas95b}.

\subsection{HD 166620: A Maunder Minimum Candidate} \label{subsec:mmcan}
HD 166620 was one of the 111 stars from \citetalias{Baliunas95b} that was classified as cycling, and estimated to have a period of about 16 yr. With the addition of new observations, the star now appears to have entered a phase of low, flat activity. 
We propose that HD 166620 is a Maunder minimum candidate. 

Unfortunately, it appears that the star's activity ``turned off" while there was a gap in data collection and a switch in our data stream from one instrument to another, between the years of 1995 and 2004. This transition into what appears to be a Maunder minimum phenomenon after change in instrumentation is very suspicious, so we undertook many checks to confirm that the change is real. 

Inspection of the observation logs at HIRES revealed no apparent errors or clues in records of this star's activity measurements. Inspection of the data for HD 166620 from HIRES show no reason that the measurements should be inaccurate. 
The data at face value argue that the grand minimum is not at a lower level than the local minima, which would have strong implications for the nature of magnetic grand minima (they are merely the end of cycling, not a shutdown of the dynamo). However, some other stars required offsets between instruments, so this conclusion is not robust.
The question is then whether Mount Wilson somehow observed some star other than HD 166620.

There are not a large number of stars bright enough and of the correct spectral type for such measurements at Mount Wilson, so a simple transcription error in the star name or routine pointing error is exceedingly unlikely. We have no ability to inspect the Mount Wilson data beyond what appears in the published tables, so we attempted to deduce the star's rough position on the sky from the median date of observations during a season. 

We use tau Ceti, HD 10700, to determine typical error in HKP measurements. The median observation date corresponds to a sidereal time at midnight of  0h52m for HKP-1, and 2h 08m for HKP-2. The real R.A. of tau Ceti is 1h44m, indicating that HKP-1 is off by 52m, and HKP-2 is off only by 23m. 

The median observation date for HD 166620 corresponds to a sidereal time at midnight of 18h39m for HKP-1, and 19h23m for HKP-2. The R.A. of HD 166620 is 18h10m. This result is very similar to that of tau Ceti, with a  difference of 29m and 1h13m for HKP-1 and HKP-2, respectively. 
Given the vagaries of seasonal weather patterns, observing strategies, and the inherent imprecision of this method, we consider that these values are all consistent with one another. We are left with the conclusion that HD 166620 suddenly switched from cycling behavior to flat in the time between the Mount Wilson and Keck surveys.


Based on the our visual analysis and curve-fitting, we estimate the cycle period for this main-sequence star prior to entering a minimum state to be 17 yr.%

With continued monitoring, we hope to capture this star through its period of minimum activity, and into its return to an activity cycle. We see evidence of a similar behavior star exiting a potential Maunder minimum period in the star HD 101501. This star, while possibly still exhibiting a low-amplitude cycle, is a good example of capturing both the drop from- and return to- cyclic behavior.  There are little recent data from CPS, but what data we have are consistent with continued cyclic variation today.

This star appears to have entered a time period of about 10 yr from 1980 to 1990 where its previously identified periodic behavior dropped to a low $S$-value with little variation. It then returned to its strong, periodic activity variation. 

When attempting to analyze this star using a periodogram, its phase of low-variation activity caused difficulty. When attempting to fit a curve with our period estimate made by eye, it became clear that the star, upon returning to its cyclic behavior, it was out of phase with its original cycle by 180. For this reason, the time series was left without a fit.

In the same manner as \citet{Shah18} with HD 4915, HIRES CPS data can be used to continue monitoring of Maunder minimum candidates as well as identify more candidates. HD 166620 shows promise for capturing the return from low, constant activity to a cycle.

\subsection{Classifications}
We have reclassified all stars that \citetalias{Baliunas95b} previously classified, and assigned new classifications to 33 stars. Some of the stars maintained their original period and classification, and some were assigned a new classification. Of the 26 stars that were included in both the \citetalias{Baliunas95b} catalog and this one, 14 stars were given new classifications based on our criteria. 

For example, HD 10700 was originally classified `flat?' and was noted to have possible increasing activity after 1988, but in the CPS data has returned to definitively flat behavior.

Of the three stars in our sample originally classified as long, only HD 141004 maintained this classification. HD 9562 and HD 143761 were reclassified as flat.


\section{Future Work}

While our work shows a significant step forward from \citetalias{Baliunas95b}, there is still a great deal of work to be done in order to reach a better understanding of stellar activity. As we continue to lengthen the time series and increase the sample size of observed stars, activity cycles will become significantly more observable, and perhaps trends will become clearer.

\subsection{Maunder Minimum Stars}
\citet{Shah18} studied the activity cycle of HD 4915 using CPS data from 2006 to 2018 and discovered a pattern suggesting the beginning of a magnetic grand minimum. Continued observations are necessary to confirm this phenomenon and to develop a better understanding of the Sun's Maunder minimum phase and magnetic fields. Our new extended time series of stellar activity will be a great resource for the continued monitoring and discovery of new Maunder minimum candidates. Stars that appear to be in Maunder minimum should be studied for coronal X-ray and chromospheric emission for comparison in the future when they return to their cycling state. This will inform studies of the dynamo and reveal whether Maunder-minimum-like events are simply extended periods of low-cycle amplitude, or periods of extraordinarily low surface magnetic field strength. 
A much more thorough search for these stars will aid in our understanding of stellar activity and the patterns, or lack thereof, in chromospheric activity.

\subsection{SSS and Other Surveys}\label{subsec:SSS}
Since the Mount Wilson HK Project came to an end, several other surveys have sparked opportunities for new research, extensions, and updates from older research, and have continued a path toward understanding of chromospheric activity.
SSS at Lowell Observatory was directly inspired by the Mount Wilson Observatory. This spectrograph incorporates both an HK spectrograph and echelle, and is an excellent resource for the observation of chromospheric activity and photospheric variability \citep{Hall07}. After the termination of the Mount Wilson survey, SSS was, and continues to be, an excellent resource to continue research on flux and chromospheric emissions. There are likely archives of unpublished data from Lowell Observatory that could provide additional data for the stars in our sample, and be added to these time series in the future.

The High Accuracy Radial velocity Planet Searcher (HARPS) spectrograph is capable of measuring high-precision Ca II H\&K indices in the same manner as the Mount Wilson survey \citep{Lovis11}. HARPS will be integral for understanding the effects of activity and jitter on radial velocity measurements and the detection of exoplanets. 

In general, more extensive and long-term observations should be done in the future. A cadence of about 10 observations per year is ideal.

\subsection{CPS Continued}

The 594 stars that were included in CPS but not the Mount Wilson surveys were not examined.  
\citet{Luhn20} noted individual stars from this sample with apparent cycles as part of their analysis of RV jitter. An in-depth analysis of the activity time series of the CPS sample is forthcoming, and will likely be the next legacy survey for activity in addition to (and in some cases in tandem with) the Mount Wilson survey.
The continued observation of these stars is important to increase the sample size allowing for concrete conclusions about what affects stellar activity. As an active collaboration studying nearby stars and planet host stars beyond
the solar neighborhood, CPS will continue to observe these stars as often as feasible in an effort
to search for planets and further understand stellar activity cycles.

\subsection{RV Follow-up Allocation}

Stars with high activity have been shown to have increased radial velocity jitter (see Section \ref{sec:intro}), which hampers the ability to detect small, Earth-like planets e.g. \citep[e.g.,][]{Saar98,Santos00,Wright05,Isaacson10}. Therefore, the best targets for RV surveys are the less active stars. It is crucial that we understand the trends between stellar properties and stellar chromospheric activity in order to build expectations for finding low-activity stars \textit{a priori}. 
It is also worth noting that some stars show strong positive correlations between activity and RVs, some do not, and some show negative correlations. Trends between activity and their translation to RVs are not clear. It has been noted that, like activity, stars experience different stages of RV jitter as they evolve \citep{Luhn20}, but that appears to vary on a star-by-star basis.


\subsection{Identification of Trends}

The availability of stellar parameters from \citet{Brewer16} allows for an examination of how stellar activity varies due to particular characteristics,
e.g., temperature, surface gravity, mass, etc. Longer-term observations could shed light on trends that were not clear based on shorter time series. It is of particular interest to understand what non-cycling main-sequence stars have in common, as well as what kind of stellar parameters cycling subgiants have. This could provide us with information deeper than how activity relates to stellar evolution. 

The relationship between effective temperature and star classification is apparent, as established by \citetalias{Baliunas95b}. \citet{Vaughan80} also identified the tendency of $\langle S \rangle$ to increase with $B-V$. Effective temperature is now available to be used in place of $B-V$ to investigate the relationship between activity and stellar evolution. 

\citet{Wright04} showed that the low-activity flat stars were all subgiants, and we see hints of this in our own analysis as well, but our sample of subgiants is small. The subgiants we did study exhibit the lowest activity levels, with consistently lower log$\langle S \rangle$ than main-sequence stars. Many of them are also classified as flat, though some show evidence of an activity cycle or variability. Analyzing the activity levels of a larger sample of subgiants could be more illuminating for this trend. The role of stellar evolution in affecting activity cycles of stars still remains unclear.

\section{Conclusion}
We have compiled five decades of curated chromospheric activity measurements, the corresponding time series and stellar parameters for 59 Sun-like stars, as well as the period estimates for all cycling stars in our sample. We have identified one Maunder minimum candidate, HD 166620, and access to stellar activity time series data that will push forward the study of stellar activity and its related topics.

\begin{longrotatetable}
\begin{deluxetable*}{lCCCrhRrrcCrcchhh}
\tablecaption{Parameters and Classifications of Program Stars\label{startable}}
\tablewidth{700pt}
\tabletypesize{\scriptsize}
\tablehead{\colhead{HD Number} & \colhead{$S_{\rm med}^{1}$} & \colhead{T$_{\rm eff}^{2}$} & \colhead{log(g)} & \colhead{$v\sin(i)$} & \nocolhead{RV Jitter$^{3}$} & \colhead{[Fe/H]} & \colhead{Mass} & \colhead{Age} & \colhead{Spectral} & \colhead{Baseline$^{4}$} & \colhead{Cadence$^{5}$} & \multicolumn{3}{c}{Classification}  & \nocolhead{Flag?} & \nocolhead{Remarks}\\
\colhead{} & \colhead{} & \colhead{(K)} & \colhead{(cm/s$^2$)} & \colhead{(km/s)} & \nocolhead{(m/s)} & \colhead{(-)} & \colhead{(M$_{\odot}$)} & \colhead{(Gyr)} & \colhead{Type$^{3}$} & \colhead{(yrs)} & \colhead{$\langle$obs/yr$\rangle$} & \colhead{Baliunas et al.1995$^{6}$} & \colhead{This Paper}$^{7}$ & \nocolhead{(Bastien)} & \colhead{} & \colhead{}}
\startdata
1388 & 0.154 & 5924 & 4.32 & 2.30 & 5.615 & 0.0300 & 1.05 & 4.3 & G0V & 27 & 10.34 & --- & flat & flat & *?-irr-see_zoom & yes(off) \\
1461 & 0.158 & 5739 & 4.34 & 1.80 & 4.023 & 0.160 & 0.970 & 4.1 & G3V & 27 & 29.44 & --- & var & flat & *see_zoom & no \\
3795 & 0.155 & 5379 & 4.11 & 1.90 & 3.612 & -0.540 & 1.67 & 11.0 & K0V & 51 & 18.11 & var & flat & flat & *irr-see_zoom & yes(off) \\
4307 & 0.145 & 5795 & 4.05 & 2.50 & --- & -0.180 & 1.12 & 7.7 & G0V & 32 & 8.33 & --- & flat & flat & *see_zoom & no \\
4628 & 0.223 & 4937 & 4.54 & 0.700 & 2.589 & -0.250 & 0.690 & 11.4 & K2.5V & 53 & 35.43 & excl-8.4 & 10.0? & cycling & *see_zoom & no \\
7924 & 0.210 & 5136 & 4.55 & 0.200 & --- & -0.160 & 0.770 & 8.7 & K0.5V & 40 & 19.49 & --- & 7.2 & cycling & *see_zoom & no \\
9562 & 0.137 & 5837 & 4.02 & 4.20 & 4.294 & 0.220 & 1.30 & 5.0 & G1V & 52 & 59.92 & long & flat & cycling & *irr-see_zoom & yes(off+diff) \\
10145 & 0.150 & 5650 & 4.43 & 1.30 & --- & -0.0400 & 1.06 & 7.3 & G5V & 40 & 2.93 & --- & flat & var & 2_cycles? & yes(off) \\
10476 & 0.186 & 5190 & 4.51 & 0.100 & 5.556 & -0.0300 & 0.780 & 8.1 & K1V & 52 & 44.86 & excl-9.6 & 10.3 & cycling & *none & no \\
10697 & 0.149 & 5600 & 3.96 & 1.50 & --- & 0.130 & 1.07 & 7.2 & G3Va & 41 & 4.64 & --- & flat & var & see_zoom & no \\
10700 & 0.169 & 5333 & 4.60 & 0.100 & --- & -0.530 & 0.990 & 12.4 & G8V & 52 & 48.24 & flat? & flat & flat & *see_zoom & no \\
10780 & 0.273 & 5344 & 4.54 & 0.800 & 4.295 & 0.0400 & 0.890 & 4.4 & K0V & 37 & 24.38 & <7 & var & var & *irr-see_zoom & yes(diff) \\
13043 & 0.149 & 5859 & 4.19 & 2.00 & 9.602 & 0.0900 & 1.07 & 5.0 & G2V & 26 & 12.39 & --- & flat & flat & *see_zoom & yes(off) \\
20165 & 0.205 & 5098 & 4.51 & 1.60 & --- & 0.0100 & 0.730 & 7.5 & K1V & 41 & 4.21 & --- & 7.8 & var & *irr & yes(2) \\
22072 & 0.131 & 4941 & 3.46 & 0.100 & 5.623 & -0.250 & 1.14 & 10.7 & K0.5IV & 52 & 18.96 & var & flat & flat & *see_zoom & yes(off) \\
23249 & 0.136 & 5037 & 3.75 & 0.100 & 4.076 & 0.160 & 1.11 & 6.7 & K0+IV & 52 & 25.58 & flat? & flat & var & *flat?-see_zoom & yes(off) \\
26965 & 0.196 & 5092 & 4.51 & 0.500 & 3.586 & -0.300 & 0.800 & 12.8 & K0V & 52 & 30.82 & excl-10.1 & 9.9 & cycling & *see_zoom & no \\
34411 & 0.146 & 5873 & 4.26 & 0.100 & --- & 0.100 & 1.08 & 4.8 & G1.5IV-V & 41 & 6.01 & --- & flat & flat & ?-see_zoom & yes(off) \\
37124 & 0.181 & 5604 & 4.60 & 0.300 & --- & -0.450 & 1.38 & 11.6 & G4IV-V & 41 & 2.49 & --- & var & flat & see_zoom & no \\
37394 & 0.451 & 5249 & 4.50 & 3.00 & 16.35 & 0.140 & 0.780 & 4.1 & K1V & 37 & 16.83 & poor-4 & var & var & *see_zoom & yes(?+diff) \\
45067 & 0.141 & 5940 & 3.92 & 5.30 & 8.210 & -0.0200 & 1.10 & 5.0 & F9V & 53 & 60.63 & flat & flat & flat & *see_zoom & yes(off) \\
50692 & 0.156 & 5913 & 4.39 & 0.100 & --- & -0.140 & 1.03 & 4.7 & G0V & 41 & 7.96 & --- & long & flat & *see_zoom & yes(2) \\
52711 & 0.158 & 5886 & 4.39 & 0.100 & --- & -0.0900 & 1.07 & 5.1 & G0V & 41 & 9.80 & --- & 11.0 & flat & long?-see_zoom & no \\
65583 & 0.167 & 5238 & 4.61 & 0.100 & --- & -0.730 & 0.930 & 13.3 & K0V & 41 & 4.38 & --- & flat & flat & see_zoom & yes(off) \\
72905 & 0.360 & 5866 & 4.50 & 9.60 & 38.64 & 0.0200 & 1.04 & 1.3 & G1.5Vb & 41 & 20.59 & var & var & var & *?-see_zoom & yes(diff) \\
75732 & 0.176 & 5250 & 4.36 & 1.70 & --- & 0.350 & 0.710 & 7.9 & G8V & 36 & 19.33 & --- & 10.9 & cycling & *see_zoom & no \\
86728 & 0.147 & 5742 & 4.31 & 2.40 & --- & 0.200 & 1.04 & 4.9 & G3Va & 41 & 6.58 & --- & flat & flat & see_zoom & no \\
100180 & 0.163 & 6002 & 4.38 & 1.70 & 5.877 & 0.0500 & 1.02 & 2.0 & F9.5V & 51 & 26.38 & fair-3.6+12.9 & var & cycling & *irr-see_zoom & no \\
101501 & 0.297 & 5502 & 4.52 & 2.20 & 13.26 & -0.0400 & 0.900 & 3.5 & G8V & 49 & 45.67 & var & 4.2? & cycling & *irr-see_zoom & yes(diff) \\
115617 & 0.164 & 5562 & 4.44 & 0.800 & 2.398 & -0.0400 & 0.930 & 7.1 & G6.5V & 41 & 44.98 & var & flat & flat & *see_zoom & no \\
126053 & 0.166 & 5714 & 4.54 & 0.100 & 4.617 & -0.350 & 1.05 & 6.5 & G1.5V & 54 & 23.47 & 22? & flat & cycling & *irr-see_zoom & no \\
132142 & 0.169 & 5145 & 4.55 & 0.400 & --- & -0.410 & 0.850 & 12.5 & K1V & 42 & 2.75 & --- & flat & flat & see_zoom & yes(off) \\
141004 & 0.156 & 5901 & 4.22 & 2.00 & 5.006 & 0.0500 & 1.15 & 4.8 & G0-V & 53 & 39.35 & long & long & flat & *see_zoom & yes(diff) \\
142267 & 0.174 & 5829 & 4.53 & 0.100 & --- & -0.410 & 1.15 & 7.5 & G1V & 35 & 3.69 & --- & flat & flat & none & no \\
143761 & 0.149 & 5833 & 4.29 & 0.100 & 3.366 & -0.210 & 1.19 & 8.4 & G0+Va & 54 & 86.60 & long & flat & flat & *see_zoom & yes(dip+diff) \\
145958A & 0.177 & 5414 & 4.48 & 1.60 & --- & -0.050 & 1.51 & 13.8 & G9V & 25 & 3.39 & --- & 7.5 & --- & --- & --- \\
145958B & 0.180 & 5343 & 4.46 & 1.70 & --- & -0.050 & 1.55 & 14.7 & G9V & 25 & 3.07 & --- & var & --- & --- & --- \\
146233 & 0.170 & 5785 & 4.41 & 1.50 & 4.594 & 0.0400 & 0.970 & 3.3 & G2Va & 39 & 25.27 & --- & var & cycling & *see_zoom & no \\
152391 & 0.386 & 5425 & 4.51 & 3.80 & --- & 0.0100 & 0.870 & 4.6 & G8.5Vk & 53 & 72.53 & excl-10.9 & 9.1 & cycling & *see_zoom & no \\
157214 & 0.155 & 5817 & 4.61 & 0.100 & --- & -0.350 & 1.79 & 8.0 & G0V & 41 & 6.67 & --- & flat & flat & ?-see_zoom & yes(dip+?) \\
159222 & 0.174 & 5870 & 4.41 & 1.00 & --- & 0.170 & 1.01 & 1.6 & G1V & 41 & 7.17 & --- & var & flat & possible_cycle-see_zoom & yes(?) \\
166620 & 0.185 & 4970 & 4.51 & 0.100 & 3.528 & -0.160 & 0.760 & 12.4 & K2V & 54 & 32.95 & excl-15.8 & 17.0? & cycling & *see_zoom & no \\
173701 & 0.193 & 5337 & 4.36 & 2.20 & --- & 0.290 & 0.720 & 5.9 & G8V & 36 & 6.33 & --- & var & var & possible_cycle-see_zoom & no \\
176377 & 0.179 & 5877 & 4.52 & 0.900 & 3.979 & -0.210 & 1.02 & 2.3 & G1V & 41 & 10.29 & --- & var & cycling & ?-see_zoom & yes(2) \\
178911B & 0.179 & 5564 & 4.40 & 2.20 & --- & 0.210 & 5.24 & 4.3 & G5D & 24 & 1.77 & --- & var & --- & --- & --- \\
179957 & 0.149 & 5741 & 4.42 & 0.100 & --- & 0.00 & 2.27 & 7.9 & G3V & 41 & 7.64 & --- & flat & flat & see_zoom & yes(off) \\
182572 & 0.148 & 5587 & 4.15 & 1.30 & 4.580 & 0.330 & 0.980 & 6.6 & G7IV & 52 & 41.26 & flat & var & cycling & *irr-peak-see_zoom & yes(off+diff) \\
185144 & 0.214 & 5242 & 4.56 & 0.500 & --- & -0.210 & 0.810 & 8.8 & K0V & 42 & 49.21 & --- & 6.2 & cycling & *decrease & yes(dip) \\
186408 & 0.150 & 5778 & 4.28 & 2.30 & 3.538 & 0.0900 & 1.03 & 5.9 & G1.5Vb & 37 & 10.37 & --- & long & var & see_zoom & no \\
186427 & 0.152 & 5747 & 4.37 & 1.60 & 3.012 & 0.0600 & 1.03 & 5.6 & G3V & 38 & 11.40 & --- & long & var & same_as_186408-see_zoom & yes(dip) \\
188512 & 0.136 & 5081 & 3.55 & 0.100 & 3.872 & -0.100 & 1.30 & 4.4 & G8IV & 51 & 25.90 & poor & var & flat & *?-see_zoom & yes(diff) \\
190360 & 0.146 & 5549 & 4.29 & 2.10 & 3.077 & 0.180 & 0.920 & 8.2 & G7IV-V & 53 & 31.48 & flat & flat & cycling & *?-low_amp-see_zoom & yes(diff) \\
190406 & 0.192 & 5940 & 4.40 & 2.30 & 7.019 & 0.0700 & 1.02 & 1.8 & G0V & 53 & 48.26 & fair-2.6+good-16.9 & 17.2? & cycling & *see_zoom & yes(2) \\
197076 & 0.179 & 5810 & 4.42 & 0.100 & 4.835 & -0.0900 & 0.920 & 3.4 & G5V & 42 & 8.40 & --- & var & flat & *see_zoom & no \\
199960 & 0.145 & 5885 & 4.22 & 2.60 & --- & 0.270 & 1.12 & 4.0 & G1V & 34 & 6.00 & --- & var & flat & irr & yes(2) \\
210277 & 0.153 & 5484 & 4.32 & 0.100 & --- & 0.160 & 0.930 & 9.6 & G8V & 38 & 6.71 & --- & flat & flat & *see_zoom & no \\
215704 & 0.248 & 5374 & 4.48 & 1.50 & --- & 0.150 & 0.840 & 3.8 & K0 & 37 & 3.39 & --- & var & var & see_zoom & no \\
217014 & 0.149 & 5758 & 4.32 & 2.00 & 2.344 & 0.190 & 1.03 & 4.7 & G2IV & 53 & 31.97 & var & flat & flat & *see_zoom & yes(diff) \\
219834B & 0.201 & 5135 & 4.48 & 1.20 & 2.278 & 0.210 & 6.37 & 5.2 & K2D & 52 & 25.86 & excl-10.0 & 9.4 & --- & --- & --- \\
\enddata
\vspace{3mm}
\tablenote{1}{Median $S$-value of star: calculated in this paper. Excludes sigma-clipped values.}\\
\tablenote{2}{\citet{Brewer16} values. Uncertainties: $T_{\rm eff} \pm$ 25K; log$(g) \pm$ 0.028; $v\sin(i) \pm$ 0.7 km/s; [Fe/H] $\pm$ 0.010. 
        Mass and age were modeled using these spectroscopic parameters. See \citet{Brewer16} for additional details.}\\
\tablenote{3}{Sourced from SIMBAD online database.}\\
\tablenote{4}{Number of years star was observed from first to final observation: calculated in this paper.}\\
\tablenote{5}{Average number of observations per year: calculated in this paper.}\\
\tablenote{6}{For the purposes of our work, we have included the FAP (false alarm probability) grade in the 
        classifications from 1995. The error in period estimate was not included, but for the stars given a 
        secondary classification, this is noted after the plus sign, e.g., HD 100180 had a primary classification 
        of period 3.6 yr and fair FAP grade, and a secondary classification of a 12.9 yr period, also given
        a fair FAP grade.}\\
\tablenote{7}{All periods listed are rough estimates, and are estimated only to the precision of one-tenth of a year. 
        This follows the same format as \citetalias{Baliunas95b} period estimates.}
\tablecomments{The final two columns compare the classifications from \citetalias{Baliunas95b}, 
                           if included in their sample, and the classifications from this work. 
                           Following a similar format as \citetalias{Baliunas95b}, our classifications for 
                           cycling stars provide the estimated period. All stars for which a periodogram analysis 
                           was not sufficient due to variable periodicity or significant gaps in data collection
                           (see Section \ref{subsec:pergram}), 
                           include a ``?” following their period estimate. These estimates were made by eye.}
\end{deluxetable*}
\end{longrotatetable}
 \label{table:startable}

\acknowledgments
{We would like to thank Dr. Fabienne Bastien for her many contributions  toward this work. 
We would like to acknowledge Olin Wilson and Sallie Baliunas for their work on stellar activity and paving the path toward this work. 
We would like to thank Arvind Gupta for his work on twilight contamination. We also thank the anonymous referee for their helpful feedback.

The Center for Exoplanets and Habitable Worlds and the Penn State Extraterrestrial Intelligence Center are supported by the Pennsylvania State 
University and its Eberly College of Science. Jacob Luhn was supported by the National Science Foundation Graduate Research Fellowship Program under Grant No. DGE1255832.
The HK\_Project\_v1995\_NSO and HK\_Project\_v2001\_NSO data derive from the Mount Wilson Observatory HK Project, which was supported by both public and private funds through the Carnegie Observatories, the Mount Wilson Institute, and the Harvard-Smithsonian Center for Astrophysics starting in 1966 and continuing for over 36 years. These data are the result of the dedicated work of O. Wilson, A. Vaughan, G. Preston, D. Duncan, S. Baliunas, and many others.

Some of the data presented herin were obtained at the W. M. Keck Observatory, which is operated as a scientific partnership among the California Institute of Technology, the University of California and the National Aeronautics and Space Administration. The Observatory was made possible by the generous financial support of the W. M. Keck Foundation. The authors wish to recognize and acknowledge the very significant cultural role and reverence that the summit of Maunakea has always had within the indigenous Hawaiian community. We are most fortunate to have the opportunity to conduct observations from this mountain.

This research has made use of
the SIMBAD database, operated at CDS, Strasbourg, France, and of NASA’s Astrophysics
Data System Bibliographic Services. 
This research made use of Astropy,\footnote{http://www.astropy.org} a community-developed core Python package for Astronomy \citep{astropy:2013, astropy:2018}.}

%

\vspace{5mm}
\facilities{}


\software{astropy \citep{astropy:2013,astropy:2018}}



\clearpage
\bibliography{references,newrefs}

\end{document}